\documentclass[a4paper,11pt]{article}
\pdfoutput=1

\usepackage{jheppub}
\usepackage{graphicx}
\usepackage{amsmath}
\usepackage{amssymb}
\usepackage{xspace}
\usepackage{caption}
\usepackage{subcaption}
\usepackage[yyyymmdd,hhmmss]{datetime}
\usepackage[normalem]{ulem}
\usepackage[export]{adjustbox}
\usepackage{xcolor}

\newcommand{\eq}[1]{eq.~\eqref{eq:#1}}
\newcommand{\eqs}[2]{eqs.~\eqref{eq:#1} and \eqref{eq:#2}}
\renewcommand{\sec}[1]{sec.~\ref{sec:#1}}

\newcommand{\app}[1]{app.~\ref{app:#1}} 

\newcommand{\fig}[1]{fig.~\ref{fig:#1}}
\newcommand{\figs}[2]{figs.~\ref{fig:#1} and \ref{fig:#2}}

\newcommand{\BI}{\begin{itemize}}
\newcommand{\EI}{\end{itemize}}
\newcommand{\BE}{\begin{equation}}
\newcommand{\EE}{\end{equation}}

\newcommand{\ord}[1]{{\mathcal O}(#1)}

\newcommand{\nn}{\nonumber}
\newcommand{\df}{\mathrm{d}}

\newcommand{\ga}{\gamma}
\newcommand{\Ga}{\Gamma}

\newcommand{\eps}{\epsilon}

\newcommand{\si}{\sigma}

\newcommand{\cI}{{\mathcal I}}

\newcommand{\GeV}{\,\mathrm{GeV}}
\newcommand{\TeV}{\,\mathrm{TeV}}
\newcommand{\gev}{\,\mathrm{GeV}}

\newcommand{\als}{\alpha_s}
\newcommand{\Gcusp}{\Gamma_{\rm cusp}^q}

\newcommand{\cut}{\mathrm{cut}}

\newcommand{\cusp}{\mathrm{cusp}}
\newcommand{\jet}{\mathrm{jet}}

\newcommand{\Ecm}{E_\mathrm{cm}}

\newcommand{\ptcut}{p_T^\cut}

\newcommand{\slepL}{\tilde{\ell}_L}
\newcommand{\slepR}{\tilde{\ell}_R}
\newcommand{\slep}{\tilde{\ell}}

\newcommand{\neuo}{\tilde \chi^0_1}

\allowdisplaybreaks[2]

\makeatletter
\newenvironment{mcases}[1][l]
 {\let\@ifnextchar\new@ifnextchar
  \left\lbrace
  \array{@{}l@{\qquad}#1@{}}}
 {\endarray\right.}
\makeatother

\title{Impact of Jet Veto Resummation on Slepton Searches}

\author[a]{Frank J. Tackmann,}
\author[b,c]{Wouter J.~Waalewijn,}
\author[c]{Lisa Zeune}

\affiliation[a]{Theory Group, Deutsches Elektronen-Synchrotron (DESY), D-22607 Hamburg, Germany}
\affiliation[b]{ITFA, University of Amsterdam, Science Park 904, 1018 XE, Amsterdam, The Netherlands}
\affiliation[c]{Nikhef, Theory Group, Science Park 105, 1098 XG, Amsterdam, The Netherlands}

\emailAdd{frank.tackmann@desy.de}
\emailAdd{wouterw@nikhef.nl}
\emailAdd{lisa.zeune@nikhef.nl}

\abstract{
Several searches for new physics at the LHC require a fixed number of signal jets,
vetoing events with additional jets from QCD radiation.
As the probed scale of new physics gets much larger than the jet-veto scale,
such jet vetoes strongly impact the QCD perturbative series, causing nontrivial
theoretical uncertainties.
We consider slepton pair production with 0 signal jets, for which we perform the resummation
of jet-veto logarithms and study its impact. Currently, the experimental exclusion limits take
the jet-veto cut into account by extrapolating to the inclusive cross section using parton shower Monte Carlos.
Our results indicate that the associated theoretical uncertainties can be large, and when taken into
account have a sizeable impact already on present exclusion limits.
This is improved by performing the resummation to higher order, which allows us to obtain accurate
predictions even for high slepton masses.
For the interpretation of the experimental results to benefit from
improved theory predictions, it would be useful for the experimental analyses
to also provide limits on the unfolded visible 0-jet cross section.
}

\preprint{
\begin{flushright}
NIKHEF 2016-010\\
DESY 16-044
\end{flushright}
}

\begin{document}

\maketitle

\section{Overview}
\label{sec:intro}

A crucial challenge at the LHC is to discriminate a faint Beyond-the-Standard Model (BSM) signal from large Standard Model (SM) backgrounds, since for most BSM searches no ``smoking gun'' signature exists. To eliminate SM backgrounds containing jets, many analyses require a fixed number of hard jets corresponding to the expected number of signal jets in the hard-interaction process. This amounts to placing a veto on additional jets above a certain transverse momentum $p_T^\cut$ arising from QCD initial-state or final-state radiation. Typical examples are supersymmetry (SUSY) searches for third generation squarks requiring two signal jets and vetoing a third jet~\cite{Aad:2013ija,CMS:2014nia,Aad:2015pfx}, or electroweakino/slepton searches requiring 0 signal jets~\cite{Aad:2014yka,Aad:2014vma,Khachatryan:2014qwa,Aad:2015jqa,Aad:2015eda}. Jet vetoes are also applied in other BSM searches, including anomalous triple-gauge couplings~\cite{Aad:2012awa}, unparticles~\cite{Khachatryan:2015bbl}, large extra dimensions and dark matter candidates in mono-photon, mono-$Z$ and mono-jet events~\cite{Aad:2012fw,Aad:2014vka,CMS:2014mea}.
In this paper, we concentrate on slepton (selectron and smuon) searches, focusing in particular on the analysis in ref.~\cite{Aad:2014vma}, which is representative of analyses with no final state jets. Searches with jets in the final state are more complicated, as the jet transverse momenta introduce additional kinematic scales in the cross section, and are left for future work.

Exclusion limits require reliable predictions for the expected BSM cross section. So far, the focus of theory calculations has mostly been on the total production cross section, while the effect of exclusive phase-space cuts like jet vetoes has not been much investigated. However, since jet vetoes impose a strong restriction on additional QCD emissions, they can significantly alter the cross section and pose an important source of theory uncertainty, as was observed some time ago in the context of Higgs production~\cite{Berger:2010xi, Stewart:2011cf}.

The jet veto introduces large logarithms in the 0-jet cross section, schematically,
\begin{align} \label{eq:vetologs}
\sigma_0(p_T^{\rm cut})
= a_{00} &+ \alpha_s \Bigl(a_{12} \ln^2 \frac{p_T^\cut}{Q} + a_{11}\ln \frac{p_T^\cut}{Q} + a_{10} \Bigr)
\nn \\ & 
+ \alpha_s^2 \Bigl(a_{24} \ln^4 \frac{p_T^\cut}{Q} + a_{23}\ln^3 \frac{p_T^\cut}{Q} + a_{22}\ln^2 \frac{p_T^\cut}{Q}
+ a_{21}\ln \frac{p_T^\cut}{Q} + a_{20} \Bigr)
\nn \\ & 
+ \dotsb
+\, (\text{terms suppressed by } p_T^\cut/Q)
\,,\end{align}
where $a_{mn}$ are coefficients and $Q$ denotes the hard-interaction scale, which is set by the (typical) partonic invariant mass, e.g.~twice the slepton mass. For $p_T^\cut \ll Q$, the logarithmic terms produce large corrections leading to a poor perturbative convergence. This can become a large effect for SUSY particle production for which $Q$ can easily be 1 TeV or more, and it will only get more important as the measurements continue to probe higher BSM scales.

The actual experimental limit is on the visible cross section in the fiducial phase space including all experimental reconstruction efficiencies and acceptance cuts, and in particular including the jet veto. Its interpretation in terms of the exclusion limits quoted by the experiments involves the extrapolation from the measured 0-jet cross section to the inclusive cross section using parton shower Monte Carlos. An important outcome of our approach is that we are able to obtain a reliable estimate of the theory uncertainty associated with the jet veto, which parton showers typically do not provide. For this reason, the jet-veto uncertainties, which we find to have a sizeable impact,  are also not taken into account in the current results that involve a jet veto.

To obtain accurate theoretical predictions and assess the theoretical uncertainties, the logarithmic terms in \eq{vetologs} can be systematically summed up to all orders in $\alpha_s$. This resummation for jet vetoes in hadronic collisions has been well-developed in the context of Drell-Yan and Higgs production~\cite{Stewart:2009yx, Stewart:2010tn, Berger:2010xi, Banfi:2012yh, Becher:2012qa, Tackmann:2012bt, Banfi:2012jm, Liu:2012sz, Liu:2013hba, Becher:2013xia, Stewart:2013faa, Banfi:2013eda, Boughezal:2013oha, Gangal:2014qda, Banfi:2015pju}, and the same methods have also been used to study diboson processes~\cite{Shao:2013uba, Li:2014ria, Moult:2014pja, Jaiswal:2014yba, Becher:2014aya, Wang:2015mvz}.

The $a_{mn}$ coefficients in \eq{vetologs} are not all independent, and their structure allows the logarithmic series to be rewritten as
\begin{align} \label{eq:vetologsresummed}
\sigma_0(p_T^{\rm cut})
&= \bigl(b_0 + b_1 \alpha_s + \dotsb\bigr)
\exp\biggl[\,\sum_{m\geq 1} \bigl(c_{0m} + c_{1m}\alpha_s + \dotsb \bigr) \alpha_s^m \ln^{m+1}\frac{p_T^\cut}{Q} \biggr]
\nn\\ & \quad
+\, (\text{terms suppressed by } p_T^\cut/Q)
\,.\end{align}
Each of the series inside round brackets is now free of logarithms, and so can be computed order by order in $\alpha_s$. Doing so then amounts to systematically performing the resummation to higher logarithmic order. The resummation orders relevant for our discussion include all terms in \eq{vetologsresummed} as follows:
\begin{align}
  \text{LL: } b_0, c_{0m}
  \,, \qquad
  \text{NLL: } b_0, c_{0m}, c_{1m}
  \,, \qquad
  \text{NLL}'\text{: } b_0, b_1, c_{0m}, c_{1m}
\,.\end{align}
The $c_1$ term, first included at NLL$'$, is important as it incorporates the full one-loop virtual corrections into the resummation, including both QCD and SUSY-QCD corrections.
The remaining terms suppressed by $p_T^\cut/Q$ in \eqs{vetologs}{vetologsresummed} start at $\ord{\alpha_s}$ and vanish as $p_T^\cut/Q \to 0$. At NLL$'+$NLO we include them at $\ord{\alpha_s}$, which then reproduces the inclusive NLO cross section in the limit $p_T^\cut\to\infty$.

\begin{figure}
\centering
\includegraphics[width=0.5\textwidth,valign=t]{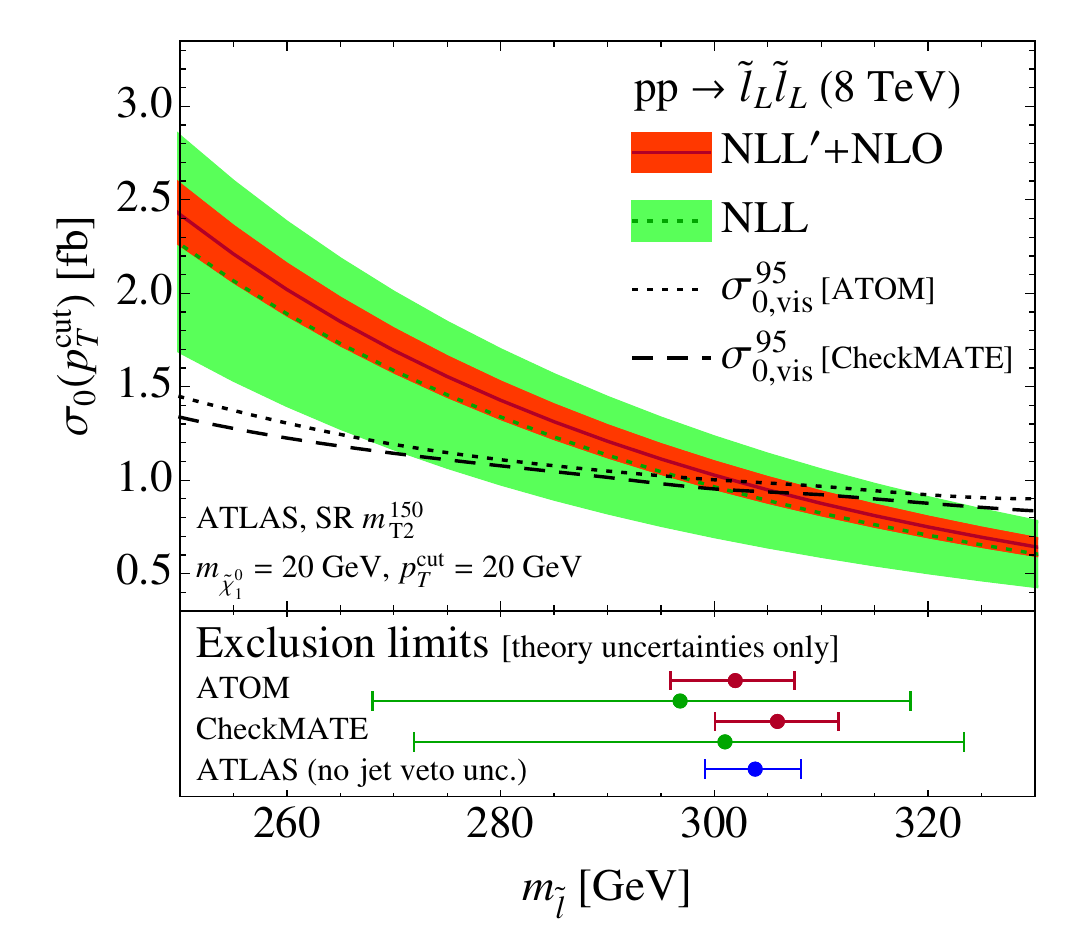}%
\hfill%
\includegraphics[width=0.5\textwidth,valign=t]{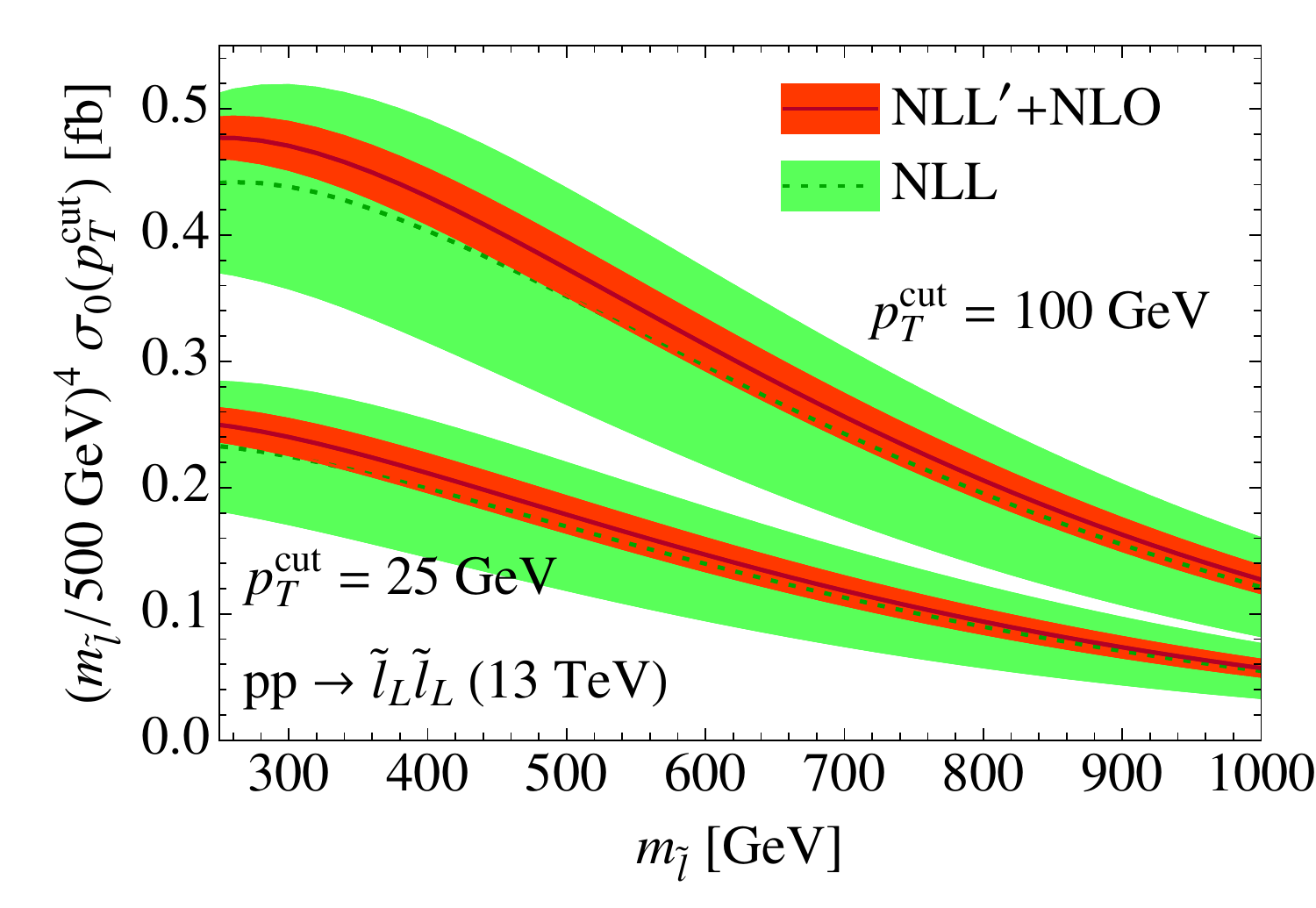}%
\caption{%
The 0-jet cross section for $\slepL \slepL$ production as a function of $m_{\slep}$ at 8 TeV (left plot) and 13 TeV (right plot).
The results at NLL are shown by the green (light) band and dotted lines and at NLL$'$+NLO by the orange (dark) band and solid lines.
In the left plot, we use $\ptcut=20 \gev$ and the dotted and dashed black lines show the experimental 95$\%$ CL upper limit on the visible 0-jet cross section, which are extracted from the ATLAS results in ref.~\cite{Aad:2014vma} using {\tt ATOM}~\cite{atom} and {\tt CheckMATE}~\cite{Drees:2013wra}. The error bars in the bottom panel give the resulting 95$\%$ CL exclusion limits on $m_{\slep}$ using our NLL prediction (green) and NLL$'$+NLO prediction (red). This is compared to the 95$\%$ CL exclusion limit provided by ATLAS (blue), which does not take into account the jet-veto uncertainty.
In the right plot, we show predictions for the $0$-jet cross section for two representative values of $\ptcut$ (25 GeV and 100 GeV), where the cross section is rescaled by a normalization factor for better visibility.}
\label{fig:summary}
\end{figure}
We now give a preview of our main results, leaving details of the calculation to \sec{resummation} and the appendices. A more extensive discussion with additional plots and results for $\slepR\slepR$ production are given in \sec{results}. Figure~\ref{fig:summary} shows our resummed predictions for the slepton production cross section with a jet veto at NLL (green band, dotted line) and at NLL$'$+NLO order (red band, solid line) as a function of the slepton mass $m_{\tilde\ell}$ for 8 TeV (left plot) and 13 TeV (right plot). In the left plot we use $\ptcut=20 \gev$, as in the  ATLAS analysis~\cite{Aad:2014vma}, and in the right plot we choose $\ptcut =25 \gev$ and $100 \gev$ as representative values. The bands show the perturbative uncertainties (but no parametric PDF uncertainties), which are systematically estimated by varying resummation and renormalization scales, as discussed in detail in \sec{unc}. The overlap between the bands and the reduction in uncertainties demonstrate the excellent stability of the resummed calculation, allowing us to obtain precise predictions even up to high slepton masses, see right panel, where the impact of the jet veto increases.

To investigate the implications for the exclusion limit, we extract the 95$\%$ CL upper limit on the visible $0$-jet cross section from the experimental results by using {\tt ATOM}~\cite{atom} and {\tt CheckMATE}~\cite{Drees:2013wra} to determine the signal region efficiencies excluding the jet veto. These are shown in the left panel as the dotted and dashed black curves. We translate this into a 95$\%$ CL exclusion limit shown as error bars in the bottom panel, using our NLL prediction (green) or NLL$'$+NLO prediction (red). This can be compared to the exclusion limit provided by ATLAS (blue)~\cite{Aad:2014vma}, for which the total NLO cross section from {\tt Prospino}~\cite{Beenakker:1999xh} was multiplied with the signal region efficiencies (including the jet veto) obtained using {\tt HERWIG++}~\cite{Bahr:2008pv}. The ATLAS exclusion accounts only for the theory uncertainty associated with the total production cross section, following ref.~\cite{Kramer:2012bx}, but does not take into account the uncertainty associated with the jet veto.

The perturbative precision of the parton shower is formally at most that of our NLL results, and hence the perturbative uncertainties due to the jet veto in the experimental limits could easily be as large as that. This has a sizeable impact: using our NLL result the exclusion would go down to $m_{\slepL}  \simeq 270 \gev$. Note that even with our NLL$'+$NLO predictions the uncertainty on the exclusion is still larger than the one obtained by ATLAS. In the future, it would be advantageous to separate out theory-sensitive acceptance cuts in the experimental results for example by quoting the observed limit on the visible $0$-jet cross section with unfolded detector efficiencies. This avoids folding a dominant theory dependence directly into the quoted exclusion limits and allows the experimental results and their interpretation to easily benefit from future improvements in theoretical predictions.

Finally, we note that soft gluon (threshold) resummation for the total slepton production cross section has been studied extensively in refs.~\cite{Fuks:2013vua, Fuks:2013lya, Broggio:2011bd, Bozzi:2007qr}. We emphasize that this type of resummation is separate and can be considered in addition to the jet veto resummation we discuss here. For current values of slepton masses under investigation at the LHC, the effect on the total cross section and uncertainty is rather small, and we therefore do not include it here.

\section{Jet veto resummation}
\label{sec:resummation}

In this section, we discuss the calculation in some detail. We utilize the jet-$p_T$ resummation of ref.~\cite{Stewart:2013faa} using soft-collinear effective theory (SCET)~\cite{Bauer:2000ew, Bauer:2000yr, Bauer:2001ct, Bauer:2001yt, Bauer:2002nz, Beneke:2002ph}.

In \sec{factorization}, we present the factorization formula for the process, $pp \rightarrow \tilde{\ell} \tilde{\ell} \rightarrow \ell \chi^0_1 \ell \chi^0_1$, and discuss how it is used to resum the jet-veto logarithms. Sec.~\ref{sec:hard} discusses the hard function that describes the underlying short-distance interaction for slepton pair production. In particular, we show that correlations between the jet veto and other kinematic selection cuts are negligible, which will allow us to ignore the slepton decay. In \sec{unc}, we explain how the theoretical uncertainties are estimated through resummation and renormalization scale variations. All fixed-order perturbative ingredients are collected in \app{FO}, while the anomalous dimensions and scale choices are summarized in \app{RGE}.

\subsection{Factorization formula}
\label{sec:factorization}

\begin{figure}
\begin{subfigure}[t]{\textwidth}
   \centering
   \includegraphics[width=0.25\textwidth]{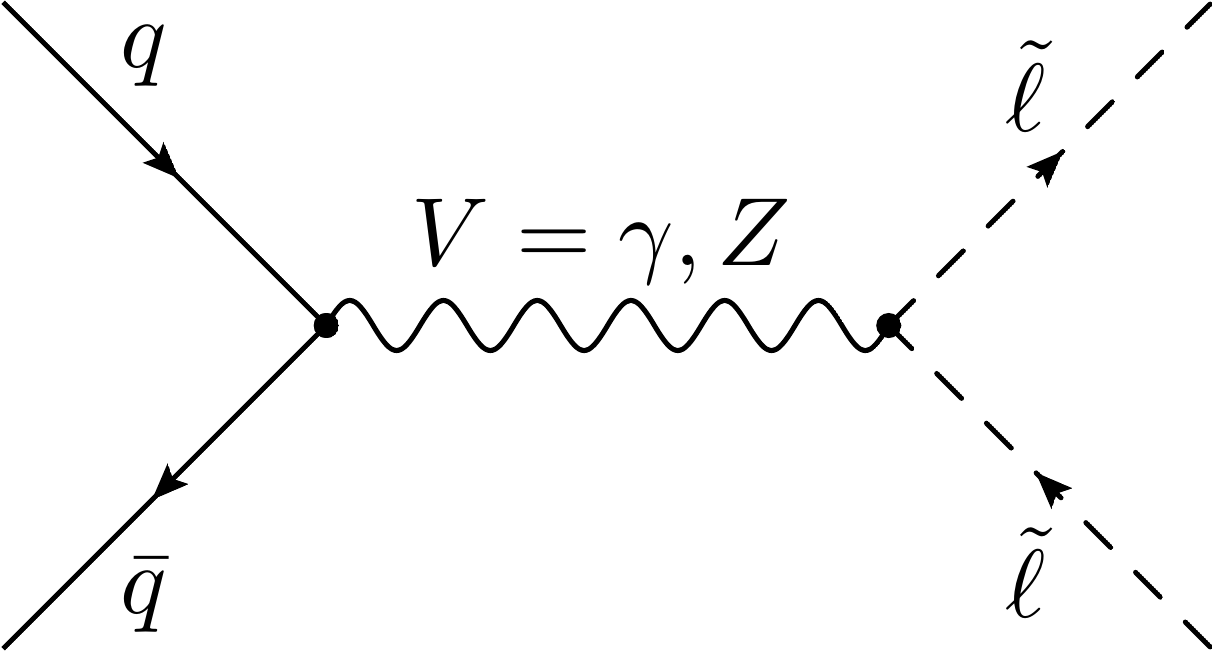}
   \caption{Leading order}
   \label{fig:LOSlepton}
  \end{subfigure}
\\[2ex]
\begin{subfigure}[t]{\textwidth}
  \centering
   \hfill
   \includegraphics[width=0.25\textwidth]{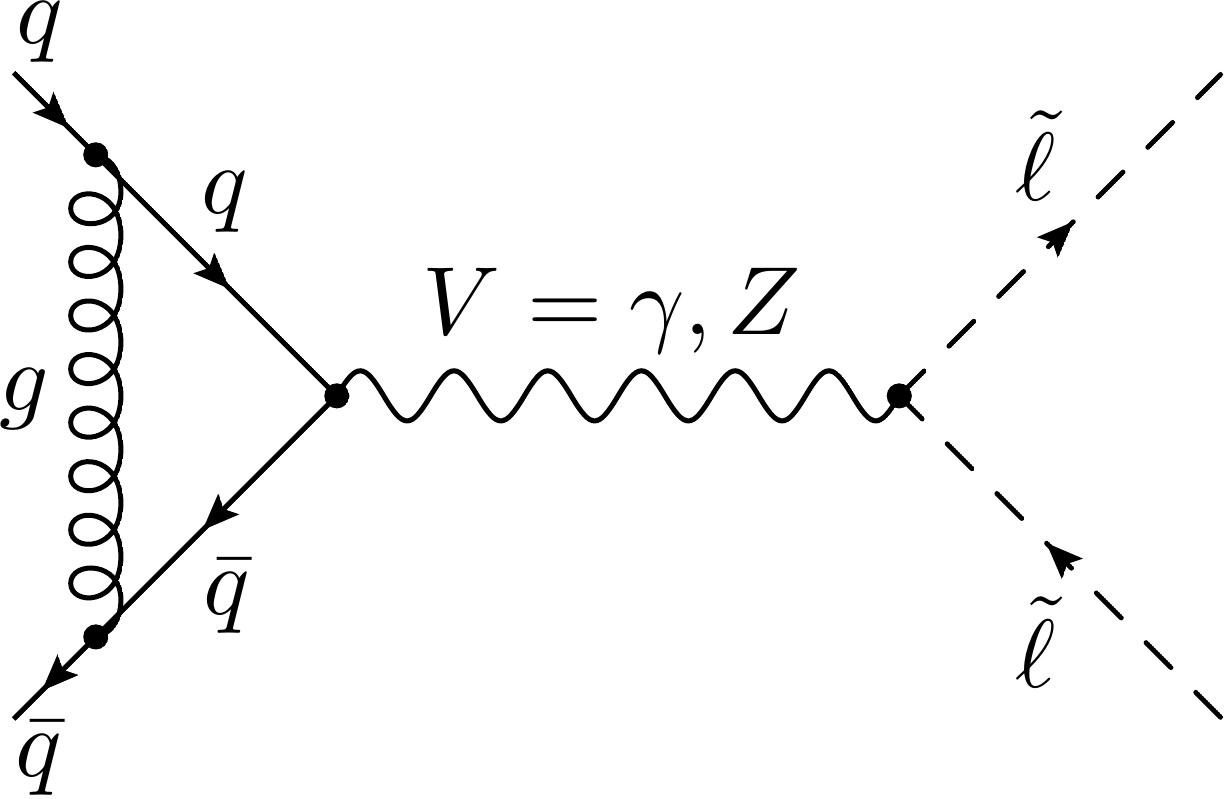}
   \hfill
   \includegraphics[width=0.25\textwidth]{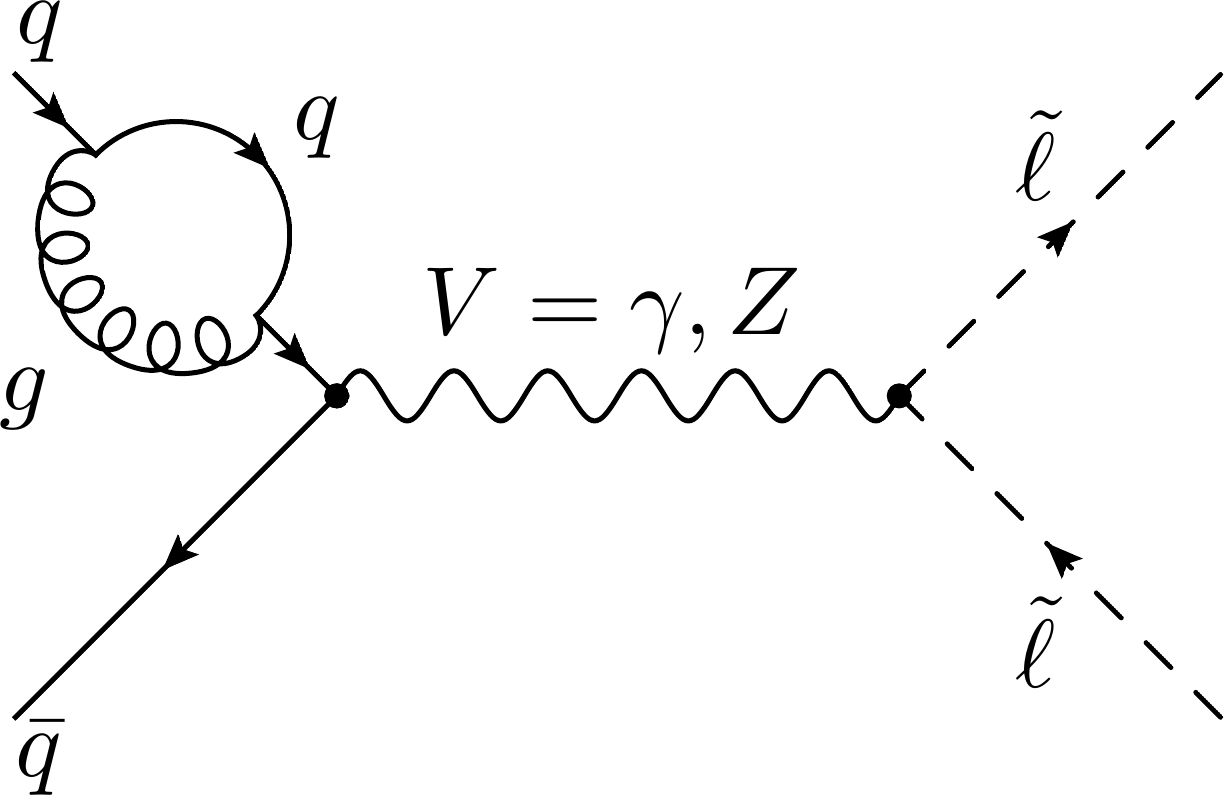}
   \hfill
   \includegraphics[width=0.25\textwidth]{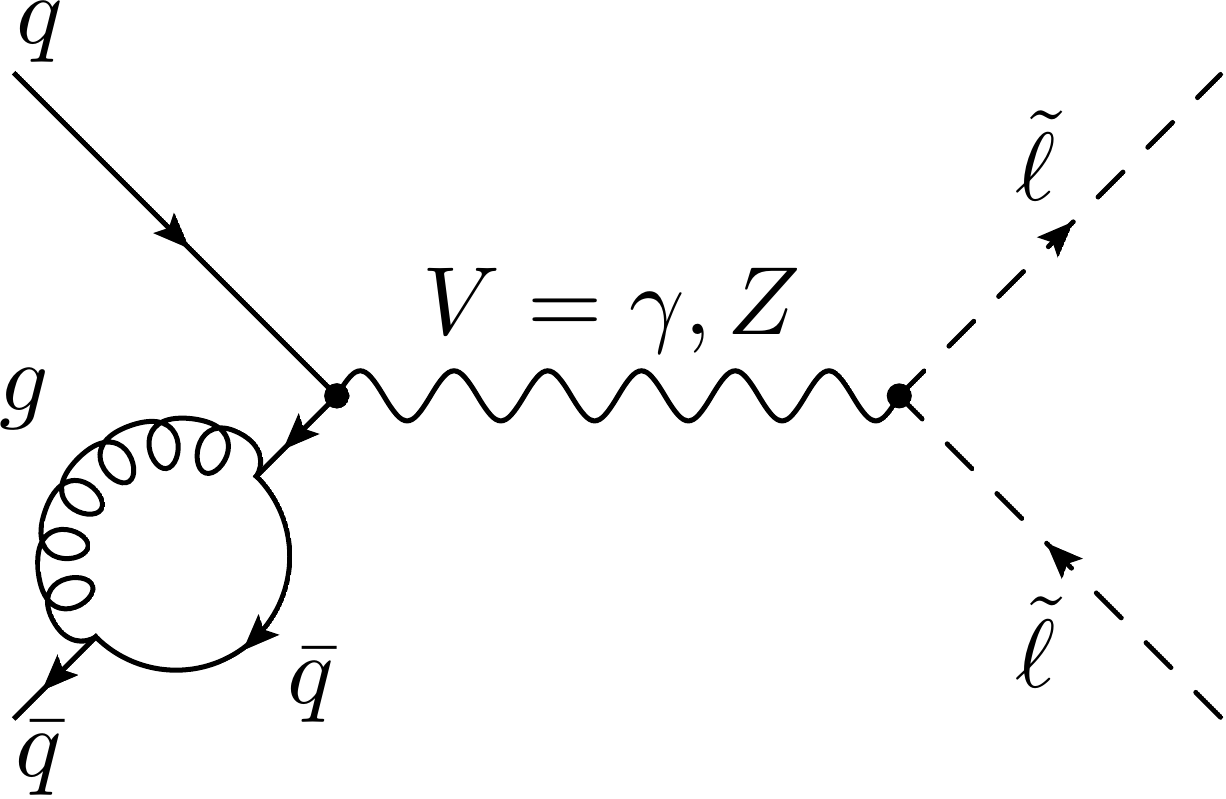}
   \hspace*{\fill}
   \caption{One-loop QCD corrections}
   \label{fig:NLOSleptonQCD}
\end{subfigure}
\\[2ex]
\begin{subfigure}[t]{\textwidth}
   \centering
   \hfill
   \includegraphics[width=0.25\textwidth]{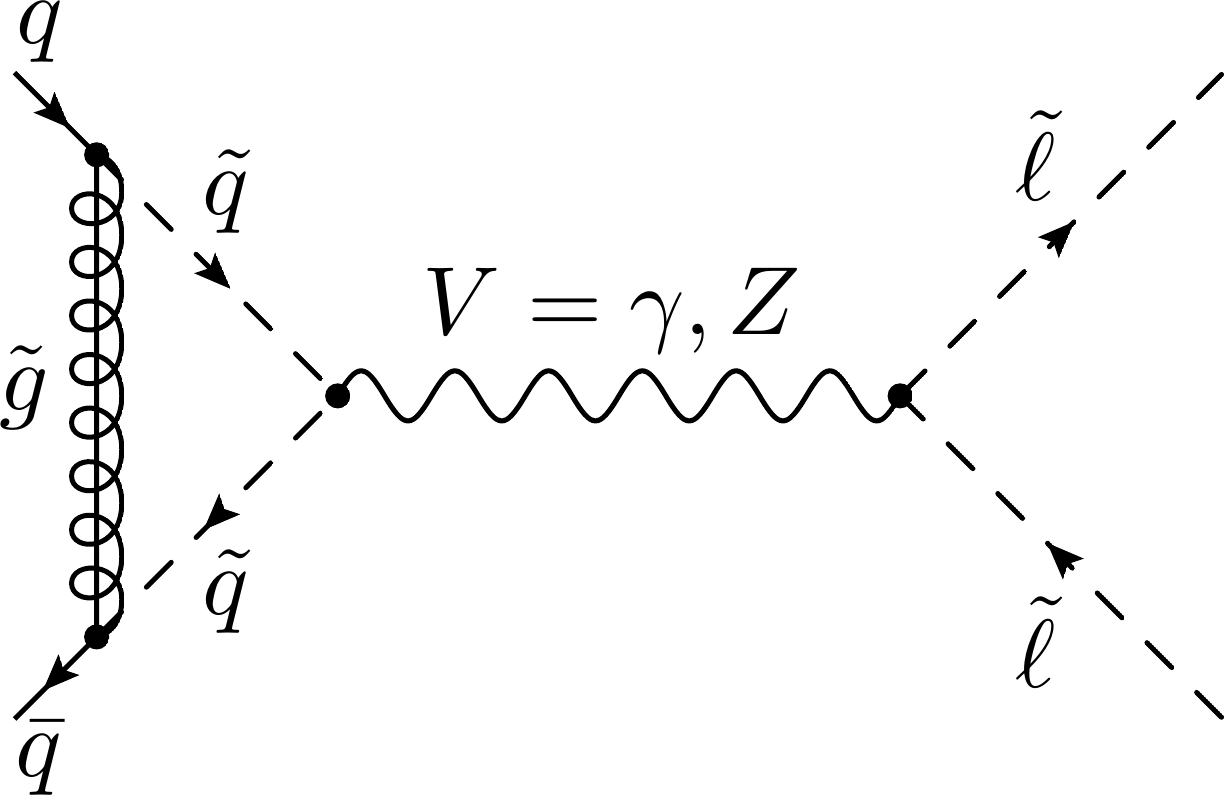}
   \hfill
   \includegraphics[width=0.25\textwidth]{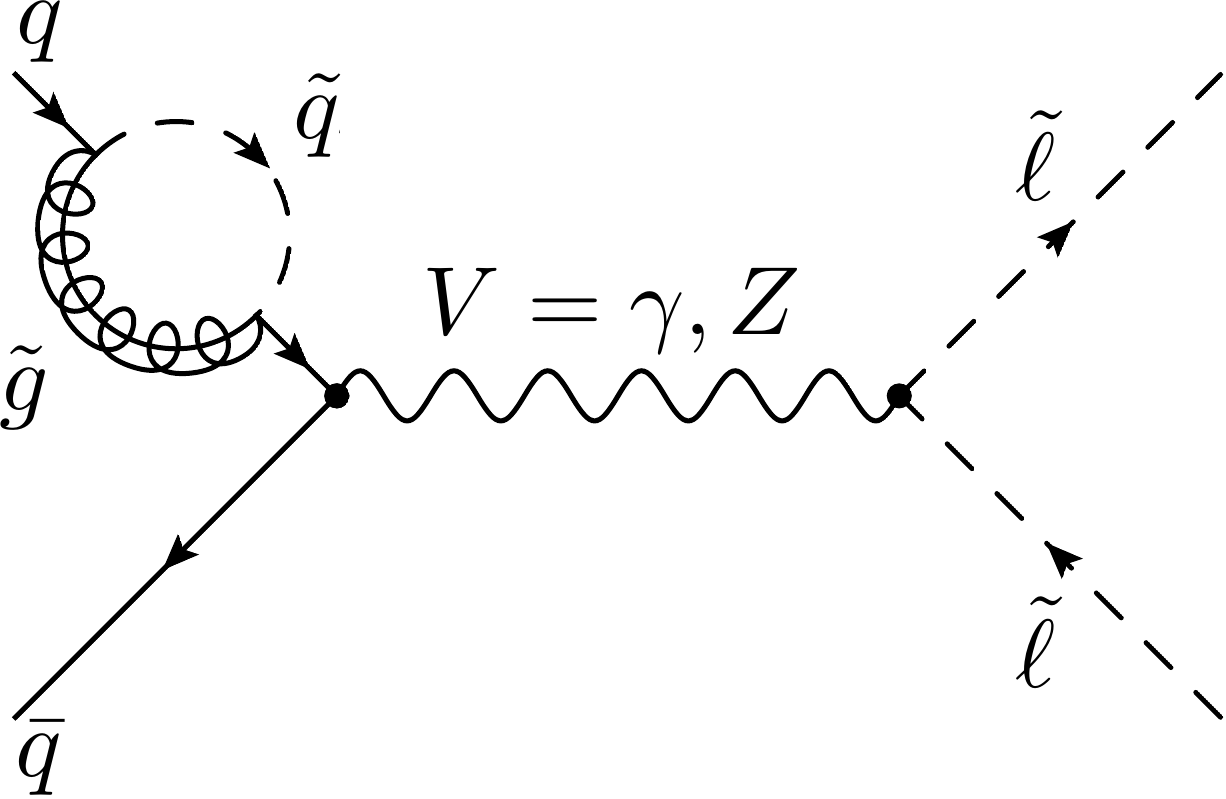}
   \hfill
   \includegraphics[width=0.25\textwidth]{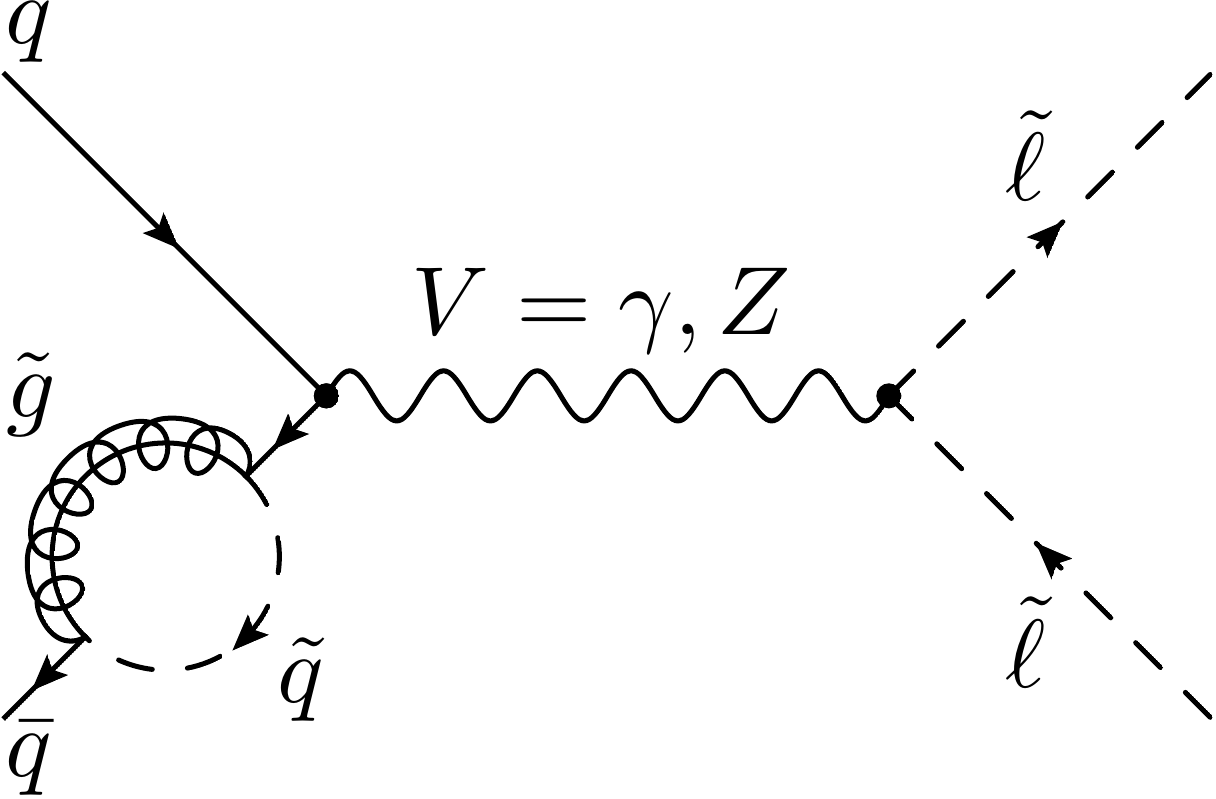}
   \hspace*{\fill}
   \caption{One-loop SUSY-QCD corrections}
   \label{fig:NLOSleptonSUSY}
\end{subfigure}
\caption{Leading order and one-loop virtual corrections to slepton pair production.}
\label{fig:Slepton}
\end{figure}

The SCET factorization formula for the $0$-jet cross section is given by~\cite{Becher:2012qa, Tackmann:2012bt}
\begin{align} \label{eq:fac}
\sigma_0(p_T^{\rm cut}, m_\text{SUSY}, \text{cuts})
&= \!\int\! \df Q^2\, \df Y\, H_{q\bar{q}} (Q^2, Y, m_\text{SUSY}, \text{cuts}, \mu)
\nn \\ & \qquad \times
B_{q} (p_T^{\rm cut}, x_a, \mu, \nu) \, B_{\bar{q}} (p_T^{\rm cut}, x_b, \mu, \nu) \, S_{q\bar q}(p_T^{\rm cut}, \mu, \nu)
\nn \\ & \quad 
+ \sigma_0^\text{nons}(p_T^\cut, m_\text{SUSY}, \text{cuts})
\,.\end{align}
Here $Q$ and $Y$ are the total invariant mass and rapidity of the sleptons, and
\begin{align}
x_a = \frac{Q}{\Ecm}\,e^{Y}
\,,\qquad
x_b = \frac{Q}{\Ecm}\,e^{-Y}
\,.\end{align}
The hard function $H_{q\bar q}$ describes the short-distance scattering process, $q \bar q \rightarrow \tilde{\ell} \tilde{\ell} \rightarrow \ell \chi^0_1 \ell \chi^0_1$. It contains all the analysis cuts applied on the slepton final state but not the jet veto.
The relevant SUSY masses are summarized by $m_\text{SUSY}$, which in addition to the slepton and neutralino masses also includes the squark and gluino masses at one-loop order (see \fig{NLOSleptonSUSY}). The hard function will be discussed in \sec{hard} and \app{hard}.

Due to the jet veto, the real QCD radiation is restricted to be collinear to the beam axis or soft. The beam function $B_q$ ($B_{\bar q}$) describes the effect of the jet veto on collinear initial-state radiation from the colliding (anti)quark with momentum fraction $x_a$ ($x_b$), and combines the nonperturbative parton distribution functions (PDFs) with perturbative initial-state radiation~\cite{Stewart:2009yx}. The restriction of the jet veto on soft radiation is encoded in the soft function $S_{q \bar q}$. The required NLO results for the beam and soft functions are given in \app{beam} and \app{soft}. The dependence on the jet algorithm and jet radius effects first appear at NNLL in $\ln(\ptcut/Q)$ and $\ord{\als^2}$~\cite{Banfi:2012jm, Becher:2012qa, Tackmann:2012bt} and are beyond the order we consider here.

The nonsingular cross section $\sigma^\text{nons}$ in \eq{fac} only consists of the $\ord{\ptcut/Q}$ suppressed terms already mentioned in \eq{vetologsresummed} and vanishes for $\ptcut\to 0$. In \app{nons} we describe how the nonsingular terms are obtained.

\begin{figure}[t]
\centering
\includegraphics[width=0.5\textwidth]{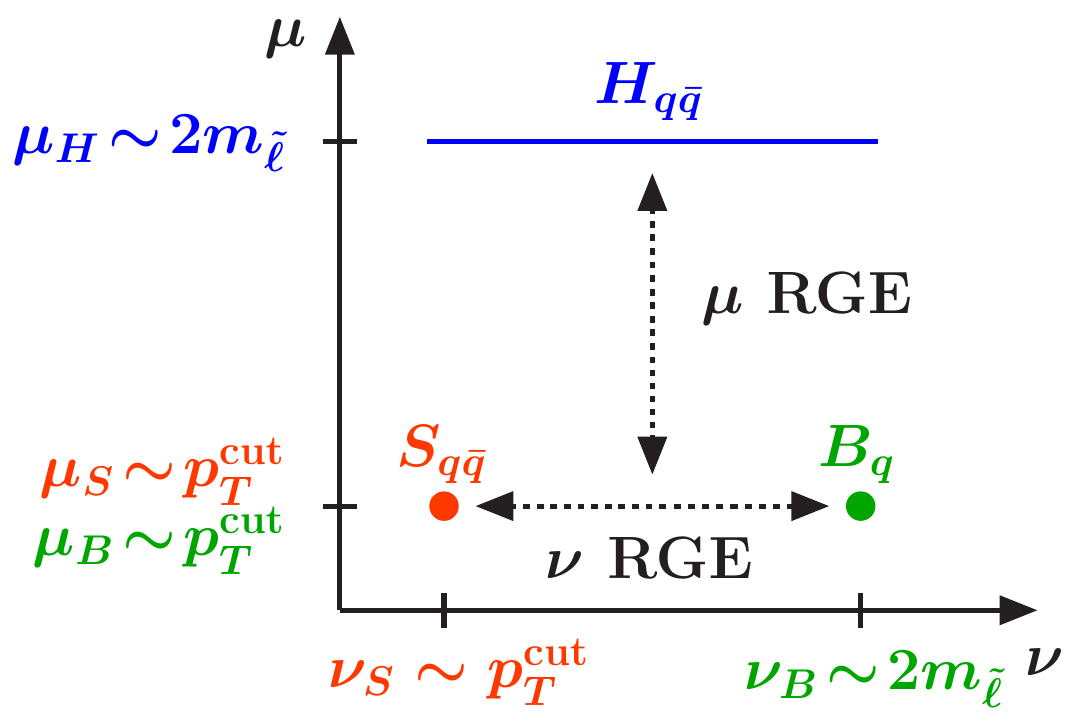}
\caption{
The hard, beam, and soft functions are evolved in virtuality $\mu$ from their natural scales $\mu_ H \sim 2m_{\tilde{\ell}}$ and $\mu_B \sim \mu_S \sim p_T^\text{cut}$. The beam and soft functions are also evolved in rapidity $\nu$ from their natural scales $\nu_B \sim 2m_{\tilde{\ell}}$ and $\nu_S \sim p_T^\text{cut}$.}
\label{fig:running}
\end{figure}

Eq.~\eqref{eq:fac} factorizes the large jet veto logarithms. For example, the leading double logarithm in the NLO cross section splits up as
\begin{align}
\ln^2 \frac{p_T^\text{cut}}{Q} = \ln^2 \frac{Q}{\mu} + 2 \ln \frac{p_T^\cut}{\mu} \ln \frac{\nu}{Q} + \ln \frac{p_T^\cut}{\mu} \ln \frac{\mu\, p_T^\cut}{\nu^2}
\,,\end{align}
where the three terms on the right-hand side are the contributions from the NLO hard, beam, and soft functions, respectively.
The key to obtaining a resummed prediction for the cross section is that each individual term can be made small by an appropriate choice of the renormalization scale $\mu$ and rapidity renormalization scale $\nu$, namely
\begin{align} \label{eq:natural}
\mu_ H \sim Q \sim  2m_{\tilde{\ell}}
\,, \qquad
\mu_B \sim \mu_S \sim p_T^\text{cut}
\,, \qquad 
\nu_B \sim Q \sim 2m_{\tilde{\ell}}
\,, \qquad
\nu_S \sim p_T^\text{cut}
\,.\end{align}
By evaluating each of the hard, beam, and soft functions at their natural scale, they contain no large logarithms. The logarithms in the cross section are then efficiently resummed by evolving each of the functions using their renormalization group evolution (RGE) for $\mu$ and the rapidity RGE for $\nu$~\cite{Chiu:2011qc, Chiu:2012ir} to the common (and arbitrary) scales $\mu$ and $\nu$ at which the cross section in \eq{fac} is evaluated. The RGE is illustrated in \fig{running} and the formulae needed for carrying it out are collected in \app{RGE}.

\begin{figure}[t]
\includegraphics[width=0.5\textwidth]{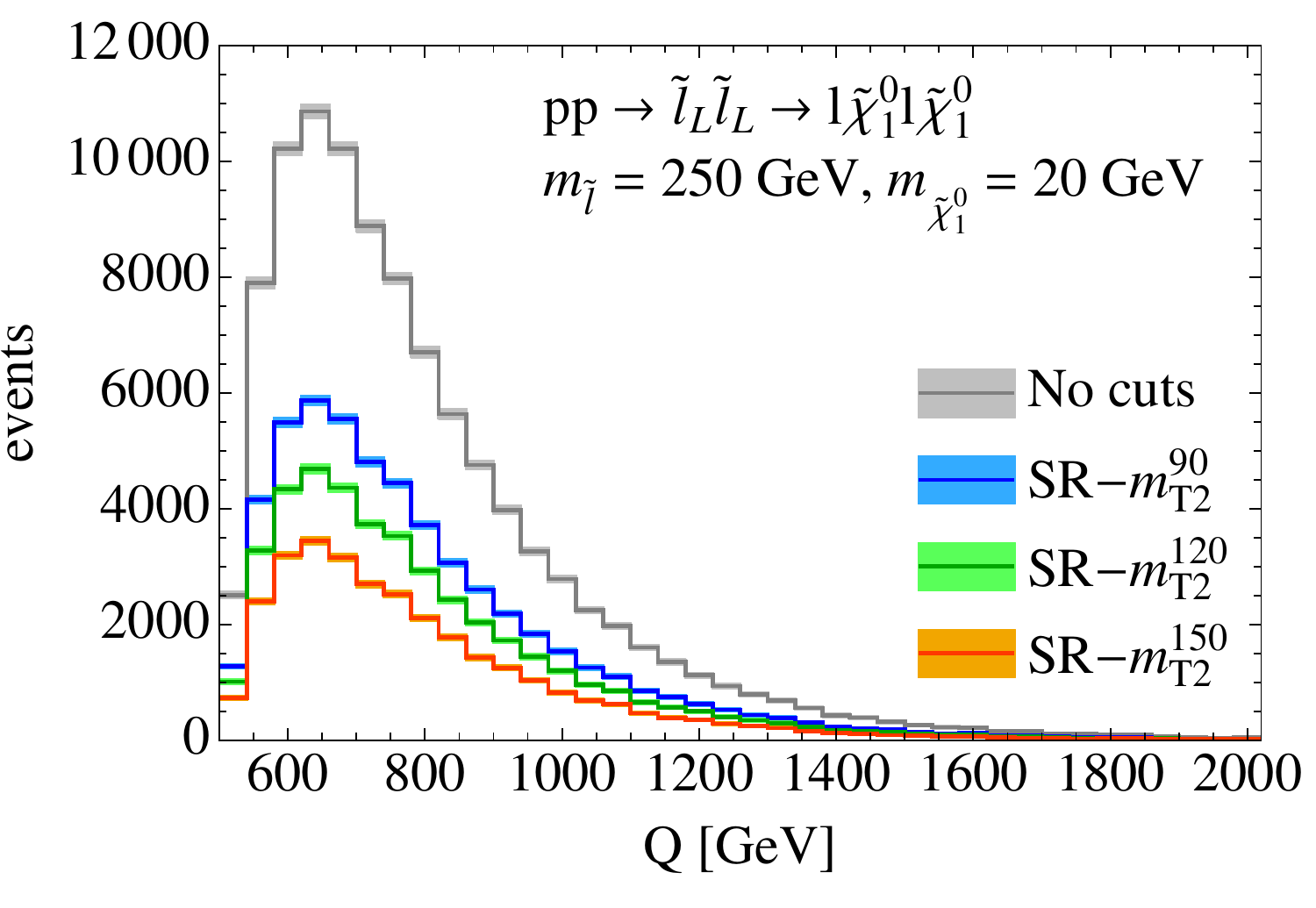}%
\hfill\includegraphics[width=0.5\textwidth]{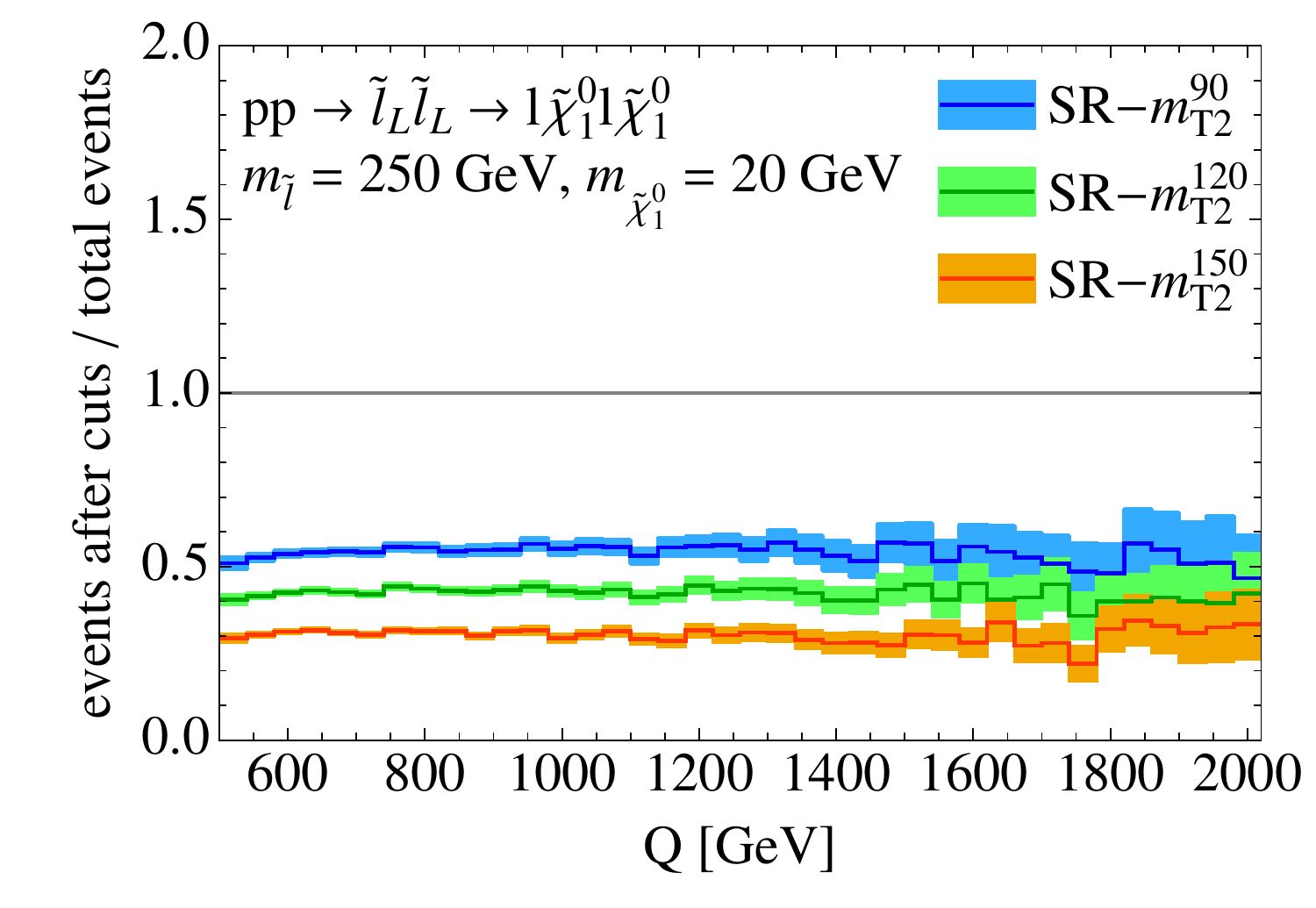}%
\\
\includegraphics[width=0.5\textwidth]{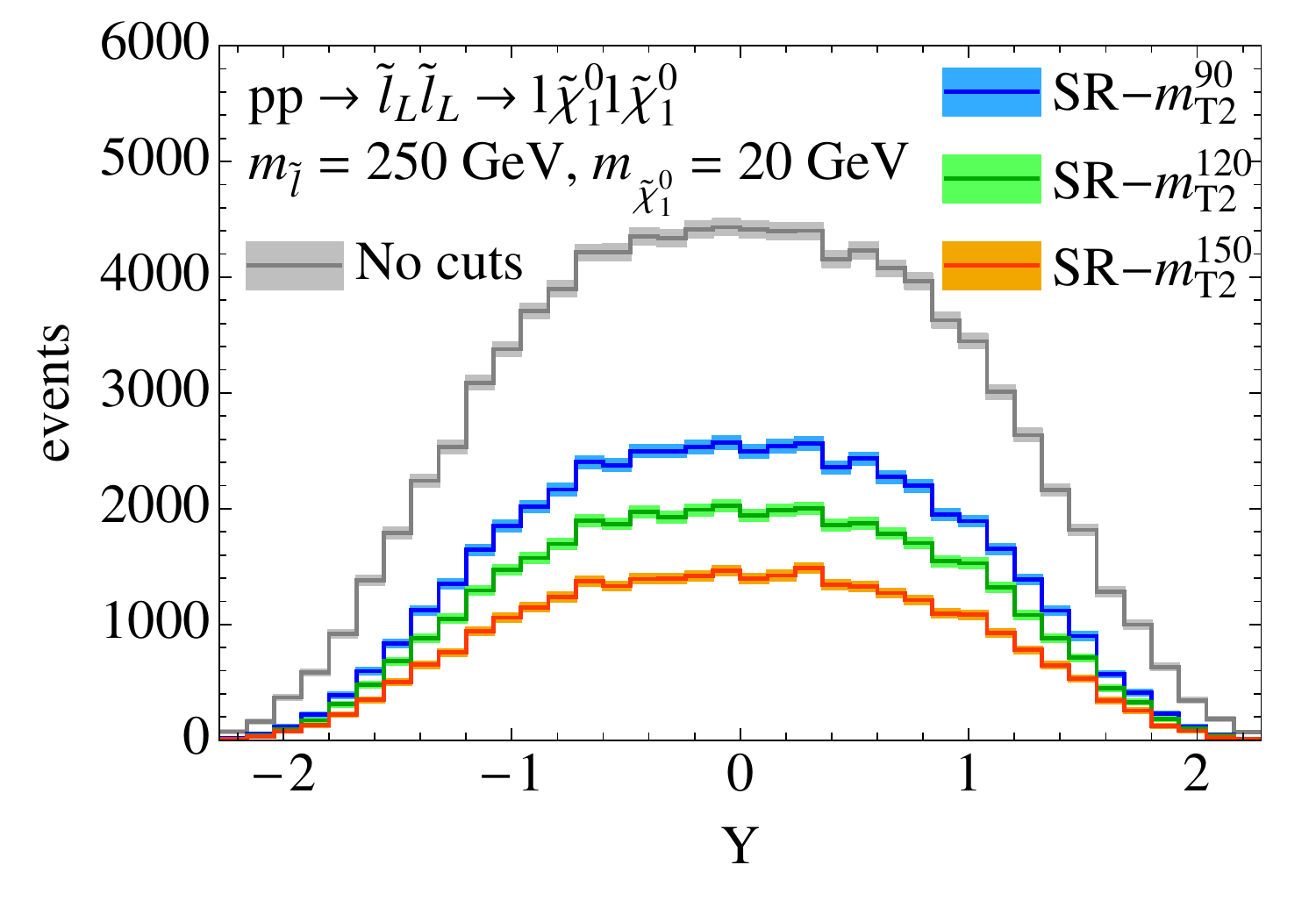}%
\includegraphics[width=0.5\textwidth]{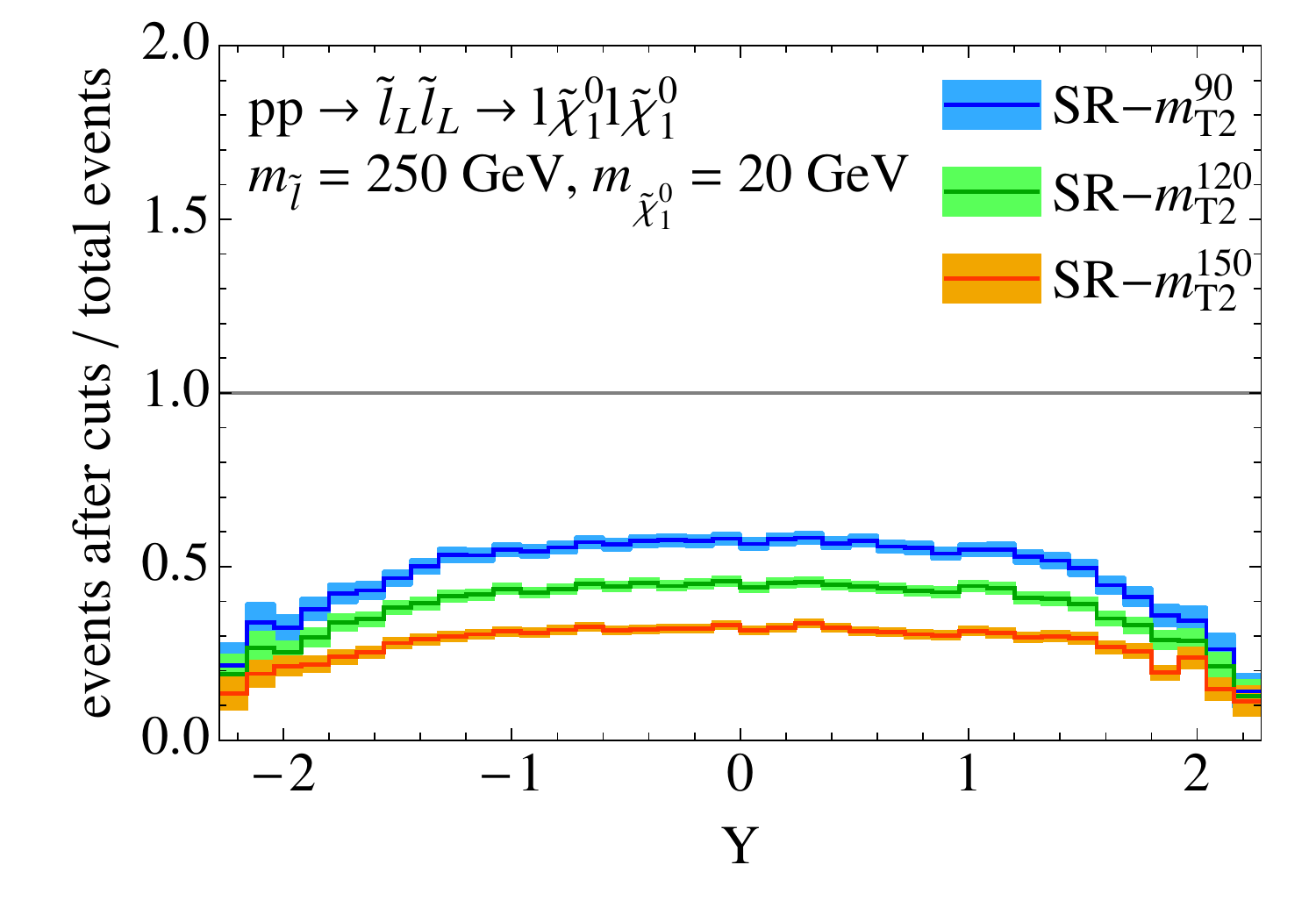}
\caption{The effect of the signal region cuts (besides the jet veto) on the $Q$ (upper row) and $Y$ (lower row) dependence of the cross section. 
The left column shows the number of events per bin, before (gray) and after (blue, green and orange) cuts. The right column shows the acceptance per bin.}
\label{fig:ratio1}
\end{figure}

\subsection{Hard scattering process}
\label{sec:hard}

We now discuss the hard function, which contains the hard scattering process $q\bar q \rightarrow \tilde{\ell} \tilde{\ell} \rightarrow \ell \chi^0_1 \ell \chi^0_1$ including the tree-level and virtual loop corrections shown in \fig{Slepton}. We consider a simplified (R-parity conserving) model where all SUSY particles except for the slepton $\tilde{\ell}$ and the lightest neutralino $\chi^0_1$ are heavy and $\mathcal{B}(\tilde{\ell} \rightarrow \ell \chi^0_1) = 1$. We will argue that we can simply calculate inclusive slepton production with a jet veto, without considering the subsequent decay of the sleptons, since the jet veto is uncorrelated with the other cuts on the slepton decay products. The resulting hard function is given in \app{hard}.

The jet veto is factorized from the other cuts in \eq{fac}, since only the soft and beam functions depend on the jet veto, whereas the hard function depends on the other cuts. Hence, the only possibility to introduce correlations between the jet veto and other cuts is through the common variables $Q^2$ and $Y$.\footnote{In principle, the hard function is independent of the boost $Y$, however the cuts are not. In addition, the nonsingular corrections $\sigma_0^{\rm nons}$ depend on both the jet veto and the other cuts, but these corrections are negligible in the relevant region of $\ptcut\ll Q$.} If the cuts were to induce sizeable changes in the $Q^2$ and $Y$ dependence of the hard function, then the $p_T^\cut$-dependent beam and soft functions would get weighted in a cut-dependent way when integrated over $Q^2$ and $Y$.

We have investigated this using {\tt MadGraph} (version 2.3.2)~\cite{Alwall:2014hca}
for the signal regions SR-$m_{T2}$ of ref.~\cite{Aad:2014vma}, which consist (besides the jet veto) of the following cuts:
\begin{itemize}
\item Two (same-flavor) leptons with $p_T > 35 \GeV$ and $p_T > 20 \GeV$. The pseudorapidity of each lepton is required to be $|\eta| < 2.47$ for electrons and $|\eta| < 2.4$ for muons.
\item The dilepton invariant mass $m_{\ell\ell} > 20 \GeV$ and $|m_{\ell\ell}  - m_Z| > 10 \GeV$.
\item Three possible cuts on the stransverse mass~\cite{Lester:1999tx,Barr:2003rg} $m_{T2} >$ 90, 120, or 150 $\GeV$.
\end{itemize}
The resulting tree-level cross section, corresponding to the tree-level hard function, is shown in \fig{ratio1} for a selectron mass of 250 GeV and a neutralino mass of 20 GeV. The gray line shows the number of events per bin without cuts and the colored lines show the number of events after the signal region cuts. The bands indicate the statistical uncertainty due to the number of simulated events. The top and bottom rows show the $Q$ and $Y$ dependence, respectively. In the right column, each bin is normalized to the total number of events in that bin, i.e., showing the acceptance of the cuts in each $Q$ and $Y$ bin. We can see that the cut acceptance is essentially flat in $Q$ and $Y$, so these cuts do not affect the shape in $Q$ and $Y$ but only the normalization. The $Y$ dependence is no longer flat for $|Y|>1.5$, but this corresponds to only 8\% of the total cross section. This implies that to very good approximation we can treat the other cuts as a $Q$ and $Y$ independent multiplicative correction which we can factor out from \eq{fac}. This treatment is completely sufficient for our purposes, since in order to compare to the experimental measurements we will also have to include experimental reconstruction efficiencies, which we are anyway only able to do approximately. Hence, we focus our attention on the jet-veto cut, which receives large QCD corrections, without considering the other cuts.

Once we restrict ourselves to only calculating the jet veto, the assumption that $\mathcal{B}(\tilde{\ell} \rightarrow \ell \chi^0_1) = 1$ allows us to focus on slepton production without the subsequent decay.
We do not consider mixing in the slepton sector and we separately discuss $\slepL \slepL$ and $\slepR \slepR$ production.%
\footnote{$\slep_{L}$ ($\slep_{R}$) denotes the
superpartner of a left-handed (right-handed) lepton $\ell$ and will be referred to as a left-handed (right-handed) slepton.}
This is a good approximation for sleptons of the first two generations, which we focus on here.
For staus, mixing effects are relevant and can be easily included.

At tree level, slepton pairs are produced via a $q\bar{q}$-initiated $s$-channel exchange of a photon $\gamma$ or a $Z$ boson, as shown in \fig{LOSlepton}. The leading-order hard function is simply equal to the corresponding partonic cross section, which has been calculated in refs.~\cite{Dawson:1983fw,Chiappetta:1985ku,delAguila:1990yw,Baer:1993ew}.
Since the intermediate $\ga/Z$ decays into a noncolored final state, the one-loop QCD corrections affect only the $q\bar{q}V$ production vertex and are identical to those of the Drell-Yan process~\cite{Altarelli:1979ub}, see \fig{NLOSleptonQCD}.
The one-loop SUSY-QCD corrections are shown in \fig{NLOSleptonSUSY}. They have been calculated in ref.~\cite{Beenakker:1999xh} neglecting squark mixing and in ref.~\cite{Bozzi:2004qq} including squark mixing. 
In the simplified model considered here, the squarks are heavy and SUSY-QCD corrections are small compared to the QCD corrections. Mixing effects in the squark sector are therefore neglected. The resulting NLO hard function is given in \app{hard}. If squark mixing effects become relevant, they can be straightforwardly included in the hard function. Note also that at one-loop order gluon-initiated slepton production is in principle also possible via a Higgs or quartic scalar coupling~\cite{Bisset:1996qh, Borzumati:2009zx}. However, the corresponding cross section is very small (except in the resonance region) and is therefore not considered here (or in {\tt Prospino}).

\subsection{Estimating the theory uncertainty}
\label{sec:unc}

In this section, we discuss the resummation scales that are used to obtain the central value for the cross section and to assess the perturbative uncertainty, with additional details relegated to \app{profiles}.
We have also evaluated the parametric PDF uncertainty for the resummed $0$-jet cross section, which is explained in the discussion of \fig{mslscannorm2} in \sec{results} below.

\begin{figure}
\centering
\includegraphics[width=0.48\textwidth]{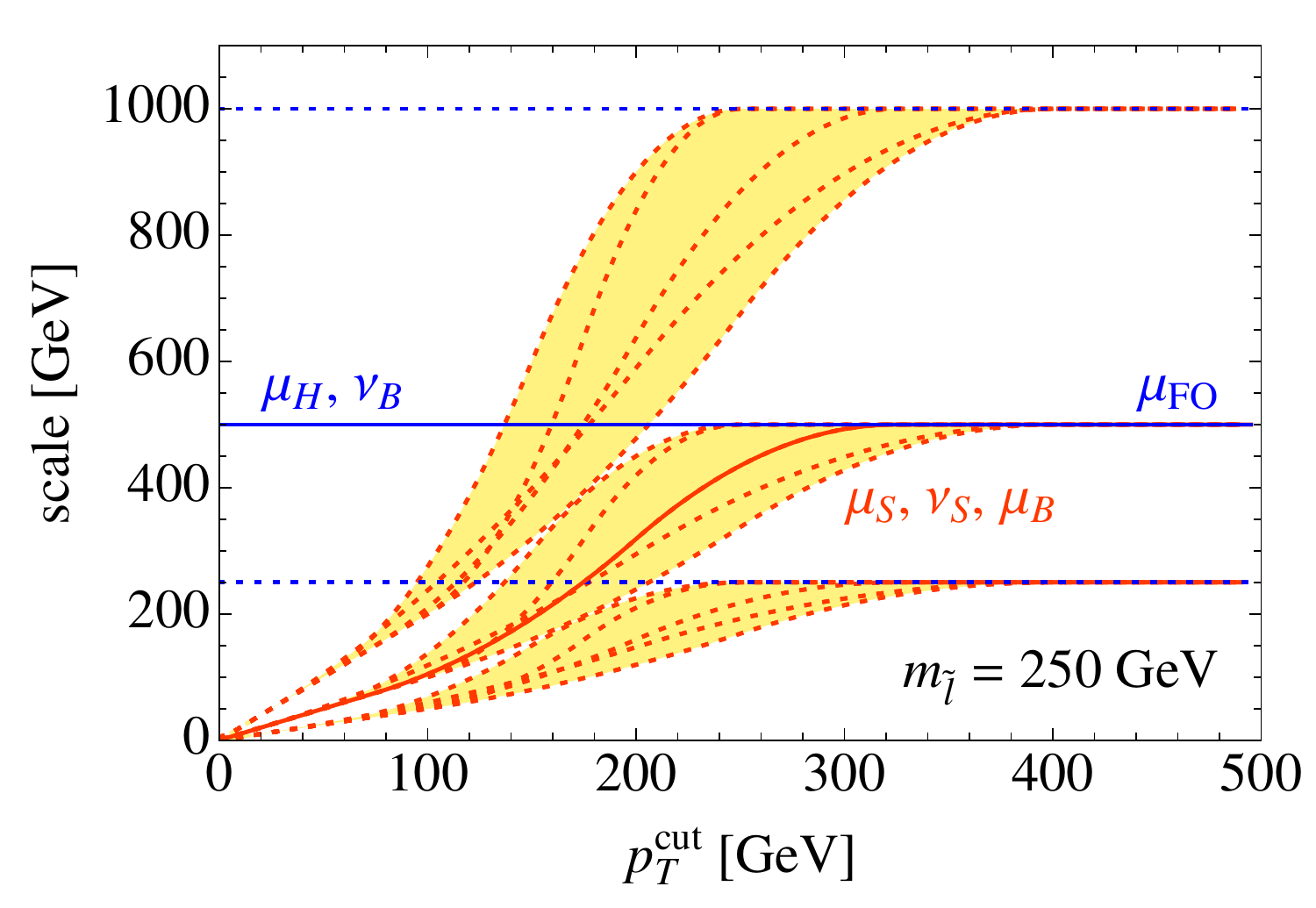}
\includegraphics[width=0.48\textwidth]{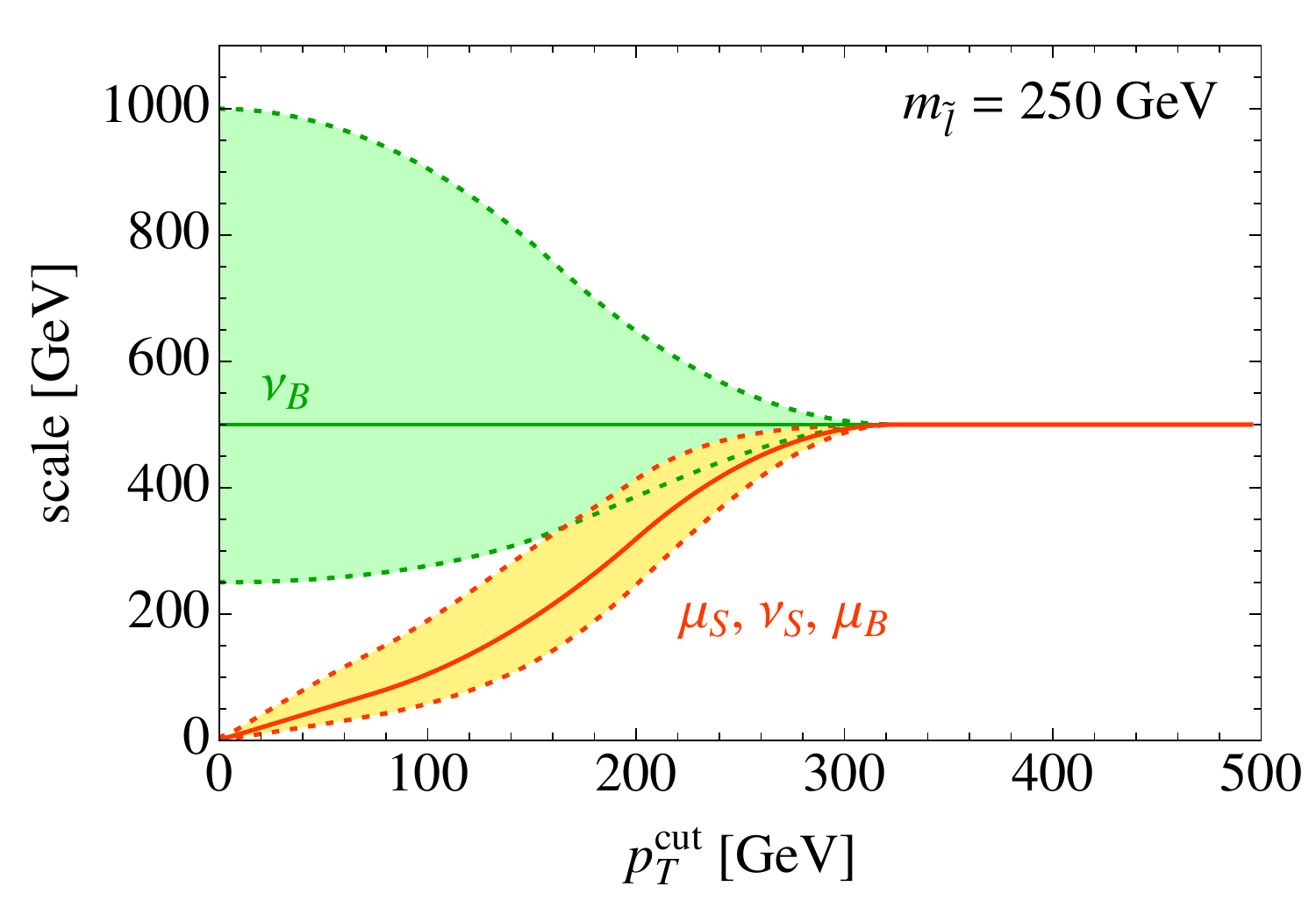}
\caption{Profile functions and their variations used to determine the theory uncertainty, as explained in the text.
Left:  Profile functions for $\mu_H, \nu_B$ in blue and for $\mu_B, \mu_S, \nu_S$ in red. 
Solid lines show the central scale choice, while dotted lines show the variations contributing to $\Delta_{\mu 0}$ where the yellow shading is between the profiles belonging to the same value of $\mu_{\rm FO}$.
Right: Variations of $\nu_B$ (green lines and shading) and $\mu_B, \mu_S, \nu_S$ (red lines and yellow shading) contributing to $\Delta_{\rm resum}$.}
\label{fig:profiles}
\end{figure}

In SCET, resummation is performed by evaluating the hard, beam, and soft functions at their natural virtuality and rapidity resummation scales and then evolving them to common $\mu$ and $\nu$ scales using their virtuality and rapidity RG equations, as illustrated in \fig{running}.
The resummation is crucial for $\ptcut \ll Q\sim 2 m_{\tilde{\ell}}$, but must be switched off for large $\ptcut$  to correctly reproduce the fixed-order cross section in that region. The smooth transition between the resummation and fixed-order regions is achieved by using $\ptcut$-depended resummation scales, called profile scales.
Profile scales were first introduced to study the $B \to X_s \gamma$ spectrum~\cite{Ligeti:2008ac} and the thrust event shape in $e^+e^-$ collisions~\cite{Abbate:2010xh}. They have since been applied in many resummed calculations and a variety of different contexts (see e.g.~refs.~\cite{Ligeti:2008ac, Abbate:2010xh, Berger:2010xi, Jain:2012uq, Jouttenus:2013hs, Stewart:2013faa, Liu:2013hba, Kang:2013nha, Kang:2013lga, Larkoski:2014uqa, Pietrulewicz:2014qza, Neill:2015roa, Alioli:2015toa, Bonvini:2015pxa,Hornig:2016ahz}) and are established as a reliable method to assess the perturbative uncertainty in resummed predictions. Our profile scales are constructed by considering the relative size of the singular and nonsingular cross section contributions, as discussed in~\app{profiles}. They are shown in \fig{profiles}, where solid curves correspond to the central scale choice and dotted curves correspond to variations that are used to estimate the perturbative uncertainty, as discussed below.

Our procedure for estimating the perturbative uncertainty using profile scale variations follows ref.~\cite{Stewart:2013faa}. The perturbative uncertainty $\Delta_0$ on the 0-jet cross section is given by
\begin{equation}
\Delta_0^2 = (\Delta_{\mu 0})^2 + \Delta_{\rm resum}^2
\,,\end{equation}
where $\Delta_{\mu 0}$ reproduces the standard fixed-order uncertainties in the limit of large $\ptcut$, whereas the resummation uncertainty $\Delta_{\rm resum}$ associated with the jet veto vanishes in the large $\ptcut$ region.
Both $\Delta_{\mu 0}$ and $\Delta_{\rm resum}$ are estimated via profile scale variations, shown in \fig{profiles}.

The set of profile variations $V_\mu$ contributing to $\Delta_{\mu 0}$ are displayed in the left panel of \fig{profiles}. They vary the overall scale by a factor $1/2$ and $2$ as well as the parameters that control the transition points between resummation and fixed-order regions. For each profile $v_i$ in $V_\mu$ we calculate the 0-jet cross section $\sigma_0^{v_i}$, from which we obtain $\Delta_{\mu 0}$ by taking the (symmetrized) envelope,
\begin{equation}
\Delta_{\mu 0} (\ptcut) = \max_{v_i \in V_\mu} \bigl\lvert \sigma_0^{v_i} (\ptcut) - \sigma_0^{\rm central} (\ptcut) \bigr\rvert
\,.\end{equation}

The profile scale variations $V_{\rm resum}$ contributing to $\Delta_{\rm resum}$ are shown in the right panel of~\fig{profiles}. They separately vary each of the beam and soft $\mu$ and $\nu$ scales up and down but keep the hard scale $\mu_H=\mu_{\rm FO}$ fixed. They thus directly probe the size of the logarithms and the associated resummation uncertainty, while smoothly turning off as the resummation itself is turned off. This yields the following estimate for $\Delta_{\rm resum}$,
\begin{equation}
\Delta_{\rm resum} (\ptcut) = \max_{v_i \in V_{\rm resum}} \bigl\lvert \sigma_0^{v_i} (\ptcut) - \sigma_0^{\rm central} (\ptcut) \bigr\rvert
\,.\end{equation}
For additional details on the profile variations we refer to \app{profiles} and ref.~\cite{Stewart:2013faa}.

\section{Results}
\label{sec:results}

In this section, we discuss our results for the $0$-jet cross section, $\sigma_0$, for slepton production
at 8 and 13 TeV and discuss the implications on current slepton exclusion limits, using the ATLAS analysis in ref.~\cite{Aad:2014vma} as a representative example.

\subsection{Slepton production at 8 TeV}
\label{sec:8tev}

\begin{figure}
\includegraphics[width=0.5\textwidth]{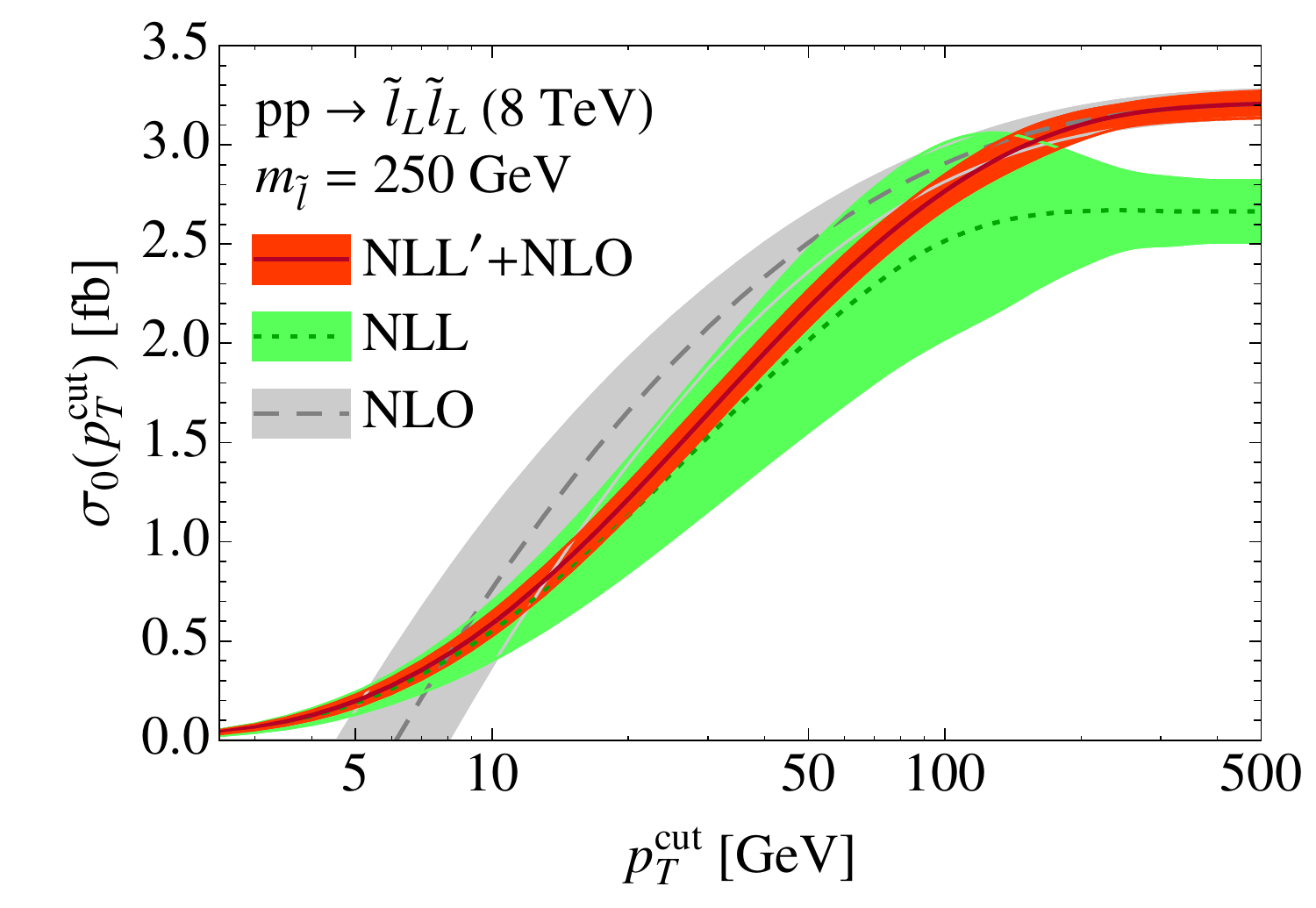}%
\hfill
\includegraphics[width=0.5\textwidth]{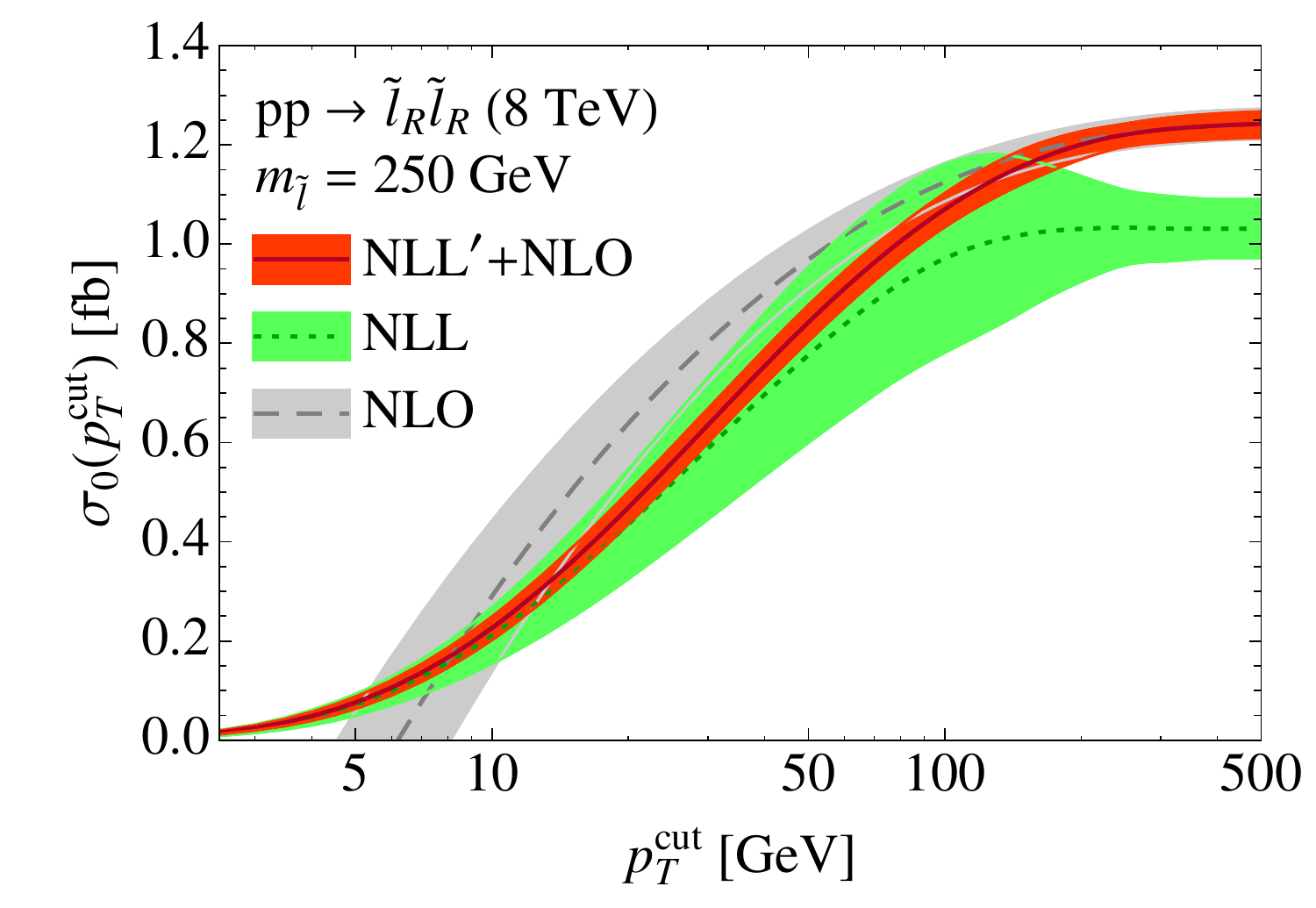}
\caption{The 0-jet cross section for $\slepL \slepL$ (left) and $\slepR \slepR$ (right) production at 8 TeV as a function of the jet veto, $\ptcut$. We compare the results at NLL (green band, dotted line), NLL$'$+NLO (orange band, solid line), and fixed NLO (gray, dashed line), where the bands show the respective perturbative uncertainties.}
\label{fig:results8}
\end{figure}

\begin{figure}
\includegraphics[width=0.5\textwidth]{Plots/8TeVMslscanPlot3.pdf}%
\hfill
\includegraphics[width=0.5\textwidth]{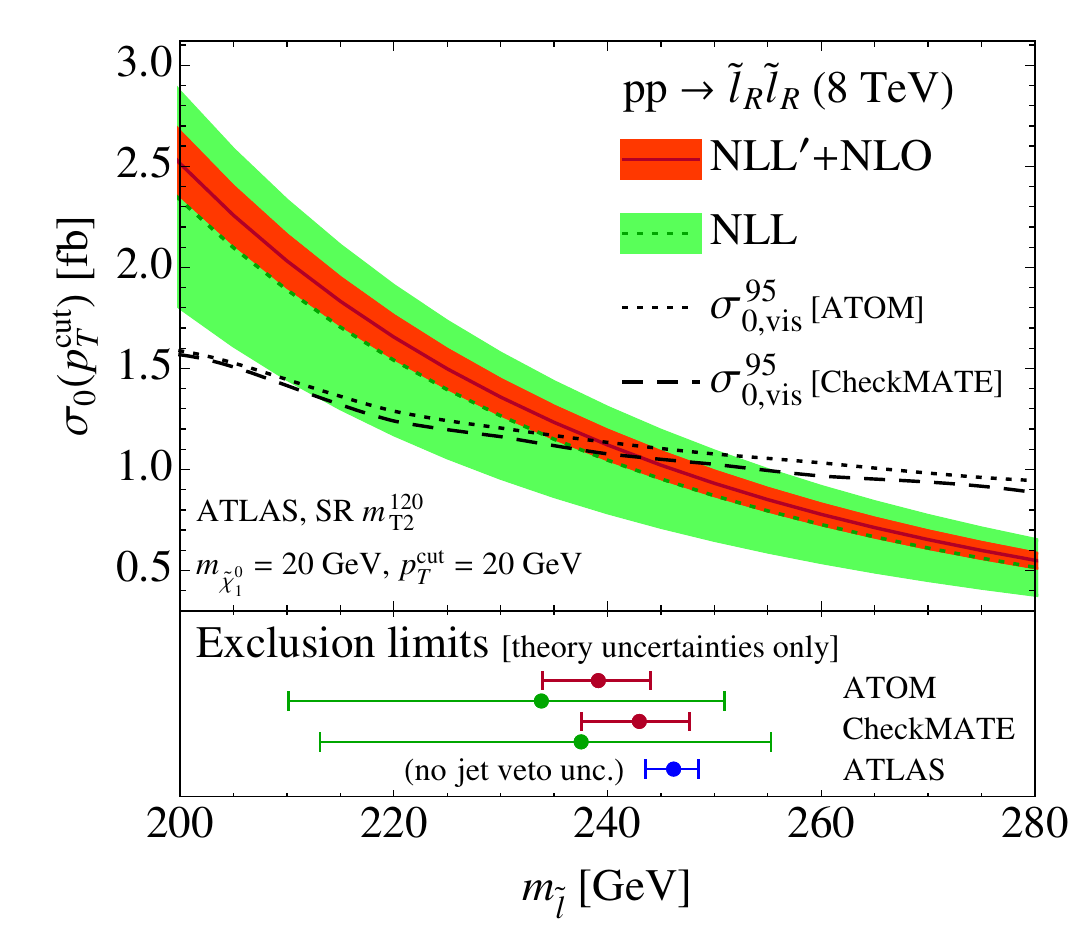}
\caption{
The 0-jet cross section for $\slepL \slepL$ (left) and $\slepR \slepR$ (right) production as a function of $m_{\slep}$ at 8 TeV. Shown are our NLL (green band, dotted line) and NLL$'$+NLO (red band, solid line) predictions, as well as the observed 95$\%$ CL upper limit on the visible 0-jet cross section, using {\tt ATOM} (black dotted line) and {\tt CheckMATE} (black dashed line) to determine the signal region efficiencies.
The error bars in the lower panels show the 95$\%$ CL exclusion limits obtained from our NLL prediction (green) and NLL$'$+NLO prediction (red), and for comparison the limit provided by ATLAS (blue).
}
\label{fig:mslscan82}
\end{figure}

We start by presenting our 8 TeV results. In \fig{results8}, we show the $p_T^\cut$ dependence of the $0$-jet cross section. This allows us to discuss the transition between the resummation and fixed-order regions, as well as the perturbative convergence and uncertainties. We consider the implications for the ATLAS exclusion limit in \fig{mslscan82}.
The {\tt CTEQ6L1} PDFs~\cite{Pumplin:2002vw} are used for the plots in this section, to remain consistent with the ATLAS analysis~\cite{Aad:2014vma}.
We show separate results for the direct production of left-handed and right-handed sleptons, focusing on the edge of the  8 TeV exclusion limits~\cite{Aad:2014vma,Khachatryan:2014qwa}.

Our predictions for the $0$-jet cross section at 8 TeV are shown in \fig{results8} as a function of $p_T^\cut$ for $\slepL \slepL$ (left panel) and $\slepR \slepR$ (right panel) production. We take $m_{\slep} = 250 \GeV$ as a representative value\footnote{For right-handed selectrons or smuons the exclusion limits are $\sim 200 \gev$, whereas for left-handed sleptons they are $\sim 275 \gev$.}, and treat other SUSY particles as decoupled. The predictions are shown at NLO (gray band, dashed line), NLL (green band, dotted line) and NLL$'$+NLO (red band, solid line). The scale choice for the central value (line) and method for estimating the perturbative uncertainty (band) were discussed in \sec{unc} and \app{profiles} for the resummed predictions.
For the NLO prediction, we use the fixed-order scale $\mu_\text{FO} = 2 m_{\slep} = 500 \gev$ for the central value and estimate the perturbative uncertainty with the ST method~\cite{Stewart:2011cf}. The latter avoids that the naive fixed-order scale variations typically underestimate the perturbative uncertainty in the fixed-order predictions for small $\ptcut$ due to cancellations between perturbative corrections to the total cross section and those related to the jet veto.

In the region $\ptcut \ll Q$, the large logarithms spoil the applicability of the fixed-order perturbative expansion and eventually drive the NLO cross section negative. A jet veto of $20 \gev$, as used in the ATLAS analysis~\cite{Aad:2014vma}, sits deep inside this resummation region. We observe that our best prediction at NLL$'$+NLO is significantly lower than the fixed NLO result.
On the other hand, fixed-order perturbation theory does provide a reliable prediction at large values of $\ptcut$, where the resummation must be turned off. Accordingly, the NLL$'$+NLO prediction smoothly merges into the NLO result, for which the nonsingular contribution to the cross section is important, as discussed in \app{nons}.
We have verified 
that in the limit of large $\ptcut$ our NLL$'$+NLO prediction exactly reproduces the NLO total cross section of {\tt Prospino}.\footnote{The default value for the fixed-order scale in {\tt Prospino} is $m_{\slep}$, which we changed to $2m_{\slep}$ for this comparison.}
Comparing the NLL and NLL$'$+NLO uncertainty bands, we find that the increased resummation and matching order leads to a substantial reduction of the uncertainties with the NLL$'$+NLO band fully inside the NLL uncertainty band (except in the fixed-order region where the uncertainties match those of the fixed-order total cross section).

Next, we investigate the implications of our resummed 0-jet slepton production cross section for the ATLAS exclusion limit~\cite{Aad:2014vma}. In their results, the visible cross section in signal region $a$  is calculated as
\begin{equation}
\sigma_{\rm vis} = \sigma (pp \rightarrow \tilde{\ell} \tilde{\ell}) \times \epsilon^{(a)}
\,,\end{equation}
where $\epsilon^{(a)}$ contains both the reconstruction efficiencies and the acceptance for the cuts of signal region $a$.
They use the total cross section $\sigma (pp \rightarrow \tilde{\ell} \tilde{\ell})$ at NLO from {\tt Prospino2.1} \cite{Beenakker:1999xh}, and determine $\epsilon^{(a)}$ using events generated by {\tt HERWIG++ v2.5.2}~\cite{Bahr:2008pv} using the {\tt CTEQ6L1} PDF set.
The resulting $\sigma_{\rm vis}$ is then compared to the measured 95\% CL upper limit on the visible BSM cross section $\sigma^{95}_{\rm vis}$ in the signal region $a$.

To compare the $\sigma^{95}_{\rm vis}$ reported by ATLAS to our predictions, we determine the upper limit on the visible \emph{0-jet cross section} as
\begin{align}
\sigma^{95}_{0,\rm vis} = \frac{\sigma^{95}_{\rm vis}}{\epsilon^{(a-{\rm no\,JV})}}
\,, \qquad
\epsilon^{(a)} = \epsilon^{(a-{\rm no\,JV})} \epsilon^{\rm JV}
\,.\end{align}
Here, $\epsilon^{(a-{\rm noJV})}$ is the signal region efficiency including reconstruction efficiencies and acceptance cuts but excluding the jet veto cut. In other words, we separate the total signal region efficiency $\epsilon^{(a)}$ into the product of $\epsilon^{(a-{\rm no\,JV})}$ and the jet veto efficiency $\epsilon^{\rm JV}$. Excluding the latter effectively avoids having to rely on the Monte Carlo to correctly describe the effect of the jet veto.
The resulting $\sigma^{95}_{0,\rm vis}$ is now defined without reconstruction efficiencies and without acceptance cuts other than the jet veto. To model the ATLAS analysis and determine the signal region efficiencies, we employ  {\tt ATOM}~\cite{atom} and {\tt CheckMATE}~\cite{Drees:2013wra}%
\footnote{Both programs use {\tt FastJet}~\cite{Cacciari:2011ma} and utilize the $m_{T2}$ variable~\cite{Lester:1999tx, Barr:2003rg, Cheng:2008hk, Bai:2012gs}. {\tt CheckMATE} applies {\tt Delphes~3}~\cite{deFavereau:2013fsa} for detector simulation, whereas {\tt ATOM} builds on {\tt RIVET}~\cite{Buckley:2010ar}. A detailed description and validation of {\tt ATOM} can be found in ref.~\cite{Papucci:2014rja}.}
Using the cut-flow tables provided by ATLAS for $m_{\slep}=250 \gev$ and $m_{\neuo}=10 \gev$, we validated both the {\tt ATOM} and {\tt CheckMATE} results for $\epsilon^{(a-{\rm noJV})}$ for the signal regions $a=m_{T2}^{120} \text{ and } m_{T2}^{150}$ of ref.~\cite{Aad:2014vma} and found agreement at the 5 - 10\% level.

Fig.~\ref{fig:mslscan82} shows the results for $\sigma^{95}_{0,\rm vis}$ as a function of the slepton mass (for a neutralino with $m_{\neuo}=20 \gev$), obtained with {\tt ATOM} (dotted black line) and {\tt CheckMATE} (dashed black line). This can be directly compared to our resummed predictions for $0$-jet slepton production at NLL (green band, dotted line) and at NLL$'$+NLO (red band, solid line).
Note that we show here the combined cross section for mass degenerate selectrons and smuons, whereas all other plots (except the left panel of \fig{summary}) are for one generation of sleptons.

The ATLAS exclusion limits were determined using the signal region with the highest expected sensitivity, which is $m_{T2}^{150}$ ($m_{T2}^{120}$) near the exclusion for left-handed (right-handed) sleptons around $m_{\slepL} \sim 300 \gev$ ($m_{\slepR} \sim 250 \gev$).
We chose these signal regions in \fig{mslscan82}, neglecting the possibility that the signal region with the highest expected sensitivity might change within the plotted range.
The intersections of the $\sigma^{95}_{0,\rm vis}$ curves with our resummed predictions set our
NLL and NLL$'$+NLO exclusion limits\footnote{We simply exclude the regions where the calculated 0-jet cross section is larger than the upper limit, $\sigma^{95}_{0,\rm vis}$, without calculating a CLs value.}, shown by the green and red error bars in the lower panels of the plots.
The blue error bars in the lower panels show for comparison the current exclusion limits as quoted by ATLAS, which account for the theory uncertainty on the total cross section (including PDF uncertainties) following ref.~\cite{Kramer:2012bx}. However, this does not include the uncertainty induced by the jet veto, which could easily be as large as our NLL uncertainty, since the perturbative precision of parton showers to model the jet veto is at best NLL.%
\footnote{Note that the Monte Carlo predictions are reweighted to the total NLO cross section. This is equivalent to rescaling the NLL green band in \fig{results8} to match the NLO result at large $\ptcut \gtrsim 2m_{\tilde\ell}$ and does not improve the resummation precision.}
At NLL the exclusion limits are noticeably weaker and would go down to $\sim 270 \gev$ for left-handed sleptons and $\sim 210 \gev$ for right-handed sleptons. Even our NLL$'+$NLO results (without including PDF$+\alpha_s$ uncertainties) yield somewhat larger uncertainties. Encouragingly, the overall central values of our best exclusion limits are similar to those obtained by ATLAS. They agree well in the left panel ($\slepL \slepL$) and are slightly lower in the right plot ($\slepR \slepR$). However, the overall central values should be treated with some caution as they rely on the signal efficiencies from {\tt ATOM} and {\tt CheckMATE}, which have 5-10\% uncertainties. To draw any firm conclusions on the final limits, the experimental analyses would need to provide results for $\sigma^{95}_{0,\rm vis}$ or to directly implement our improved theoretical predictions and uncertainties in their interpretations.

\subsection{Slepton production at 13 TeV}
\label{sec:13tev}
\begin{figure}
\includegraphics[width=0.5\textwidth]{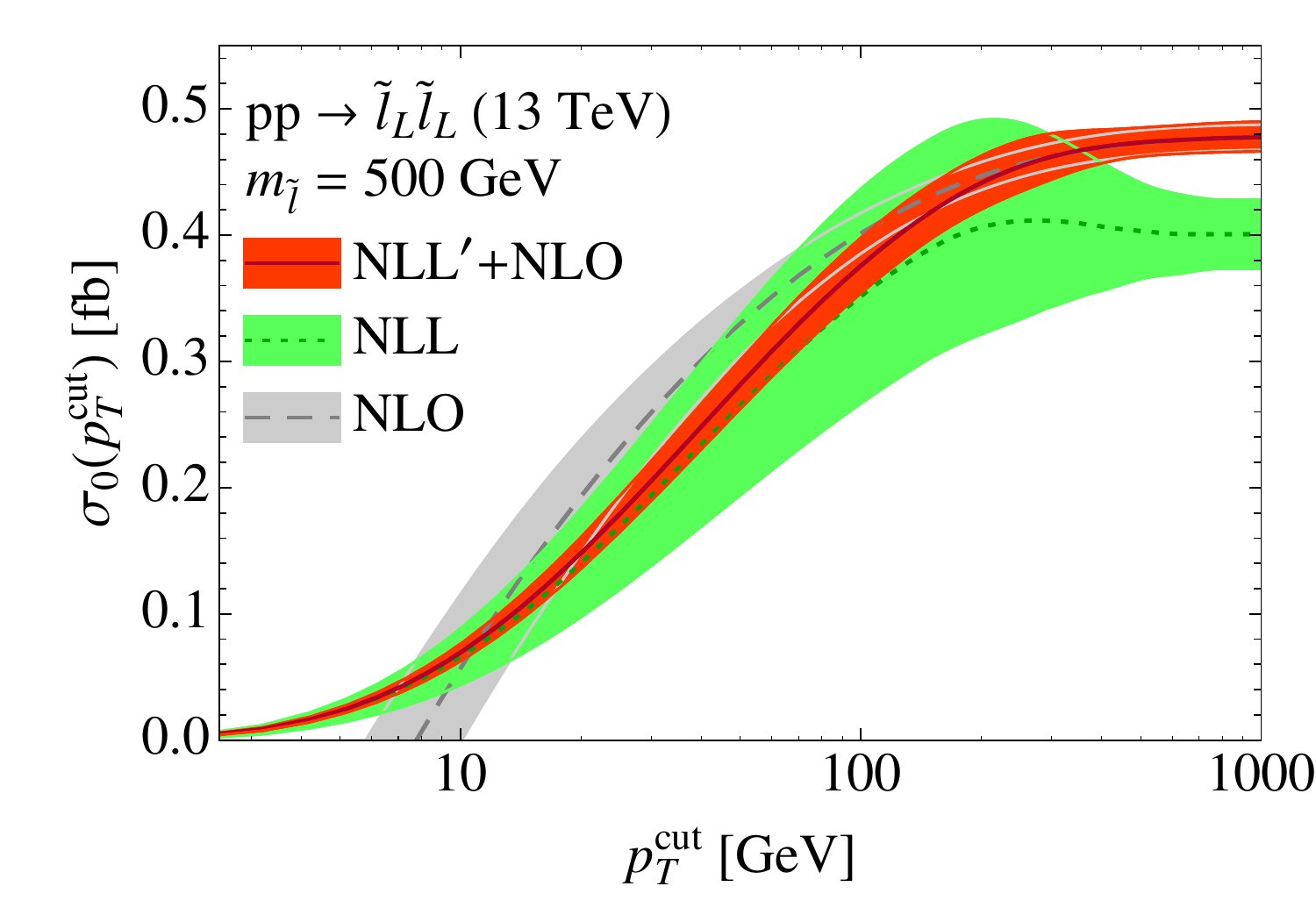}%
\hfill
\includegraphics[width=0.5\textwidth]{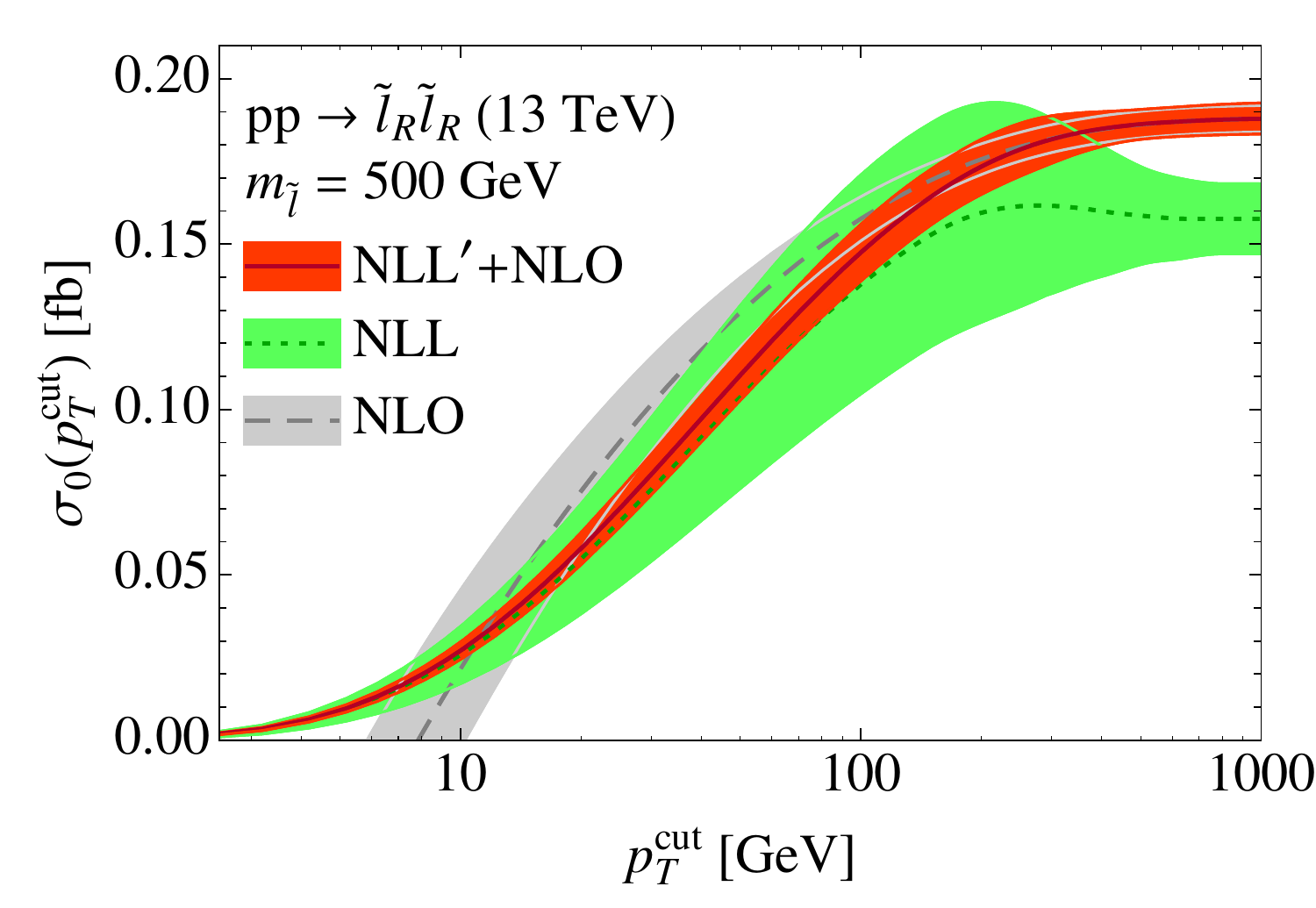}
\caption{
The 0-jet cross section for $\slepL \slepL$ (left) and $\slepR \slepR$ (right) production as a function of $\ptcut$ for $m_{\slep}=500 \gev$ at 13 TeV. The bands show the perturbative uncertainties.}
\label{fig:results13}
\end{figure}

\begin{figure}
\includegraphics[width=0.5\textwidth]{Plots/13TeVMslscan_LLPlot2.pdf}%
\hfill
\includegraphics[width=0.5\textwidth]{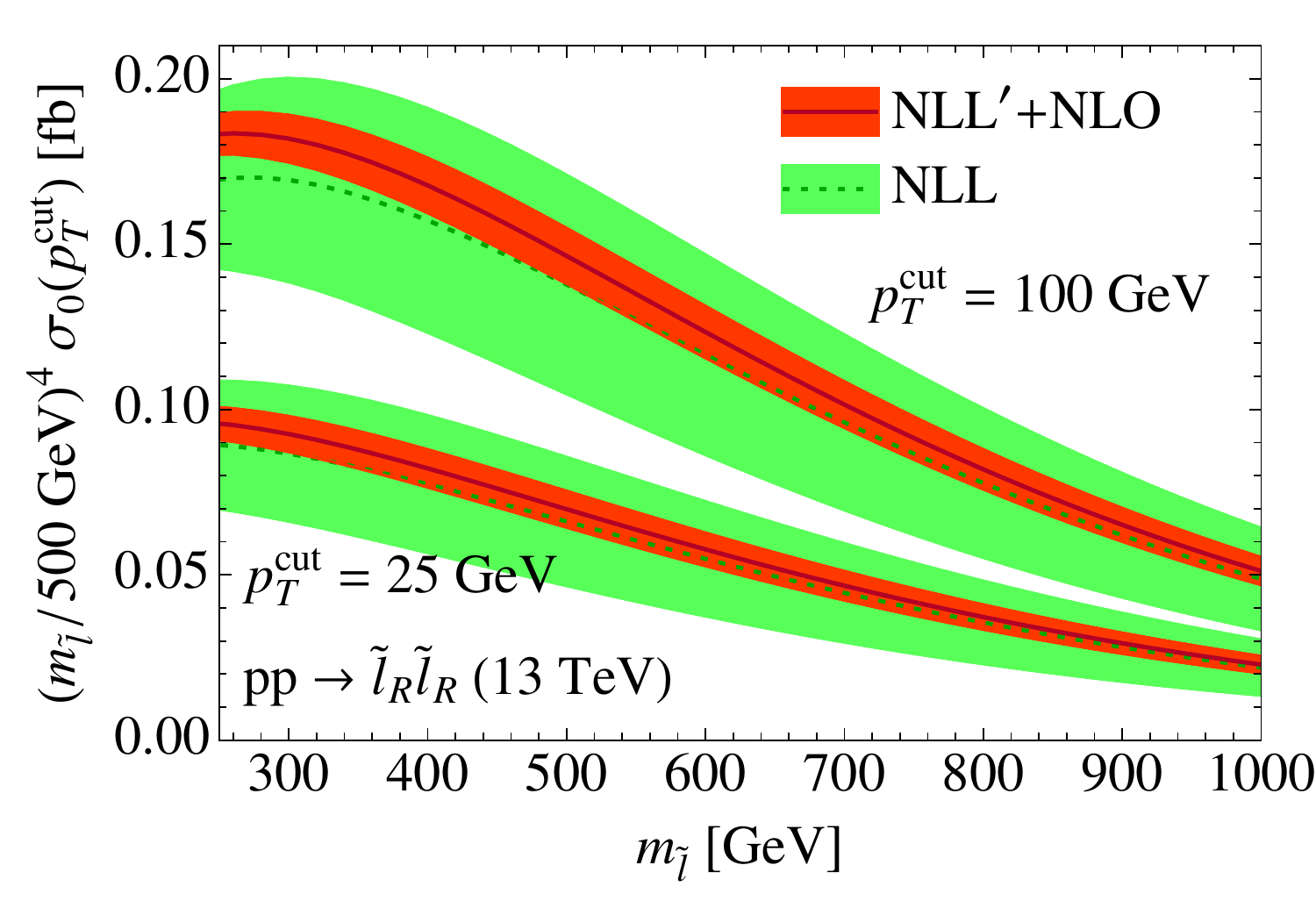}
\caption{
The 0-jet cross section for $\slepL \slepL$ (left) and $\slepR \slepR$ (right) production as a function of $m_{\slep}$ for $p_T^\cut = 25\gev$ and $100 \gev$ at 13 TeV. Shown are the NLL (green band, dotted line) and NLL$'$+NLO (red band, solid line) predictions with their perturbative uncertainty. We multiply the cross section by $(m_{\tilde \ell}/500 \gev)^4$ for better visibility.}
\label{fig:mslscannorm1}
\end{figure}

\begin{figure}
\includegraphics[width=0.5\textwidth]{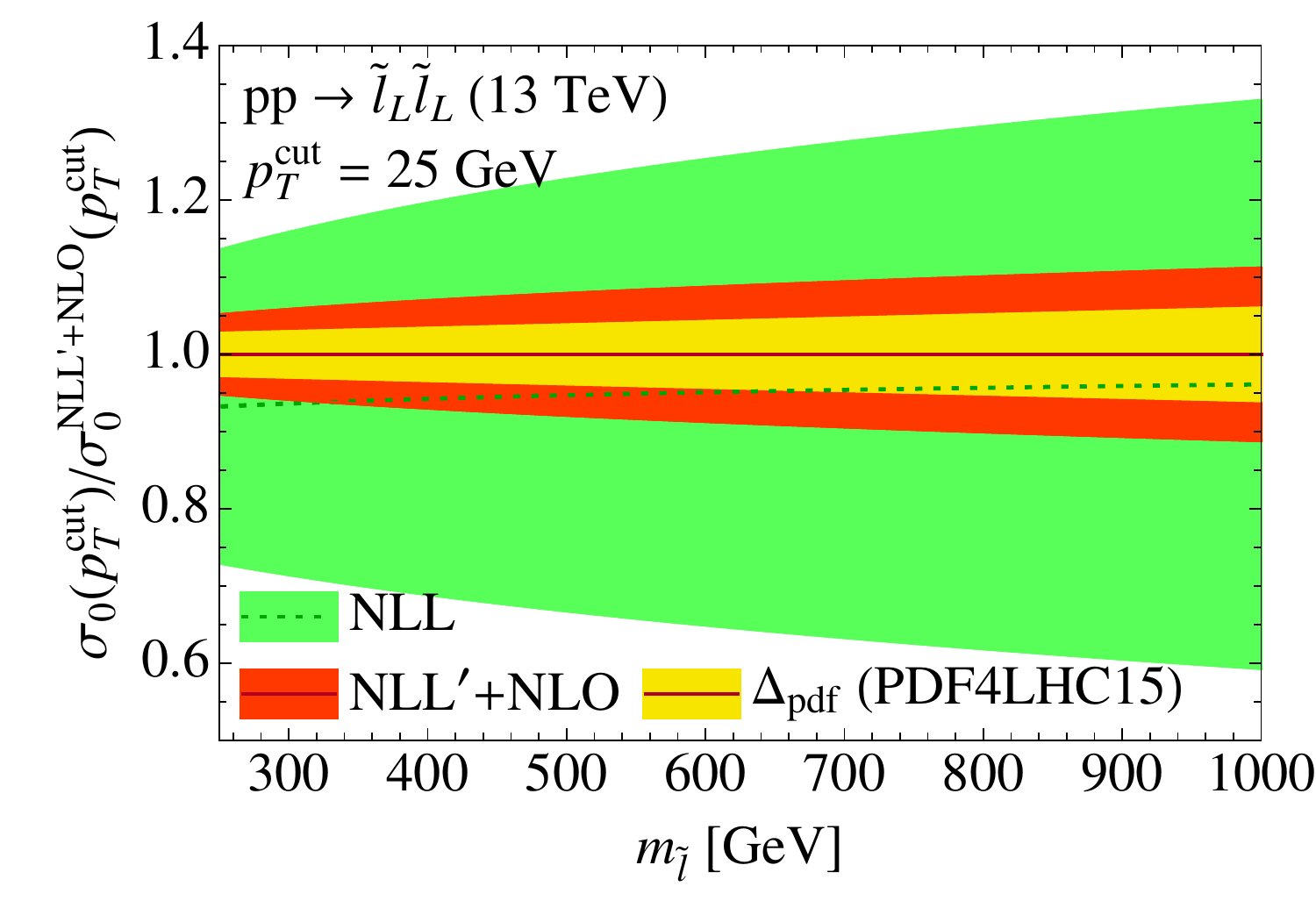}%
\hfill
\includegraphics[width=0.5\textwidth]{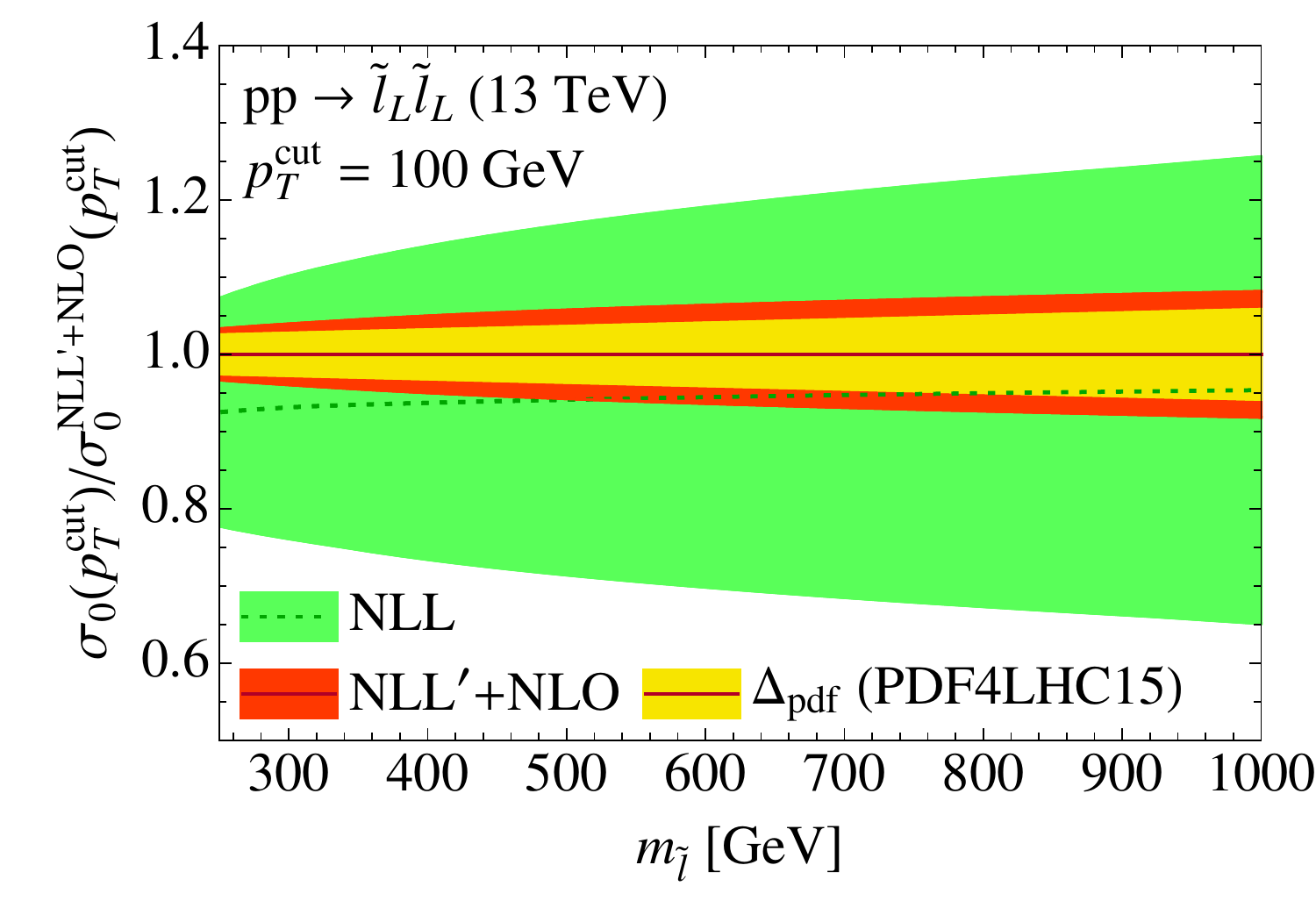}
\caption{The 0-jet cross section for $\slepL \slepL$ as a function of $m_{\slep}$ for $p_T^\cut = 25\gev$ (left) and $100 \gev$ (right) at 13 TeV. The predictions are normalized to the NLL$'$+NLO central value. The NLL and NLL$'$+NLO perturbative uncertainty are shown by the green and orange band, respectively. The yellow band shows in addition the PDF uncertainty for the NLL$'$+NLO results, determined following ref.~\cite{Butterworth:2015oua}.}
\label{fig:mslscannorm2}
\end{figure}

We continue our discussion with the 0-jet cross section for slepton production at 13 TeV.
Following the PDF4LHC recommendations~\cite{Butterworth:2015oua}, we use the {\tt PDF4LHC15\_nlo\_mc} PDF set in this section.
In \fig{results13}, we show our resummed results for the 0-jet cross section as a function of $p_T^\cut$ for $m_{\slep}=500 \gev$. Comparing this to the 8 TeV results with $m_{\slep}=250 \gev$ in \fig{results8}, we observe an increase in the perturbative uncertainties. This is expected due to the higher slepton mass, which leads to larger logarithms in the cross section.

In \fig{mslscannorm1}, we show the resummed 0-jet cross section as a function of $m_{\slep}$ for $\ptcut = 25\gev$ and $100 \gev$.
The nonsingular contribution is small enough that we can neglect it in this plot.\footnote{Even for $\ptcut=100 \gev$ and $m_{\slep}=250 \gev$ where it is least suppressed the nonsingular correction is only $\sim 1\%$. For larger $m_{\slep}$ and smaller $\ptcut$ the nonsingular contribution is significantly smaller.}
The overlap between the NLL and NLL$'$+NLO bands illustrates again the excellent stability of our resummed calculation.

In \fig{mslscannorm2} we focus on the uncertainties, normalizing all results to the central NLL$'$+NLO result. The 0-jet cross section for left-handed slepton production is shown for $\ptcut=25 \gev$ (left panel) and $\ptcut=100 \gev$ (right panel) at NLL (green band, dotted lines) and NLL$'$+NLO (red band, solid lines). Furthermore, the yellow band shows the PDF uncertainty of the NLL$'$+NLO result, obtained using the standard deviation approach in ref.~\cite{Butterworth:2015oua}.%
\footnote{An alternative method to calculate the PDF uncertainties is given in eq.~(24) of ref.~\cite{Butterworth:2015oua}. Here the uncertainty is determined by reordering the cross sections obtained from the member PDFs and taking the spread between 68\% most central ones, which is particularly suitable when the departure from the Gaussian regime is sizeable. We have checked that this method leads to slightly smaller uncertainties in our case. E.g.~for $\ptcut=25 \gev$ and $m_{\slep}=600 \gev$, the PDF uncertainty obtained from the standard deviation is $4.3\%$, whereas the PDF uncertainty calculated with the reordering method is $4.0\%$.}
The perturbative uncertainty is still larger than the PDF uncertainty, so we are not yet limited by the latter, though they become comparable for $\ptcut=100 \gev$.
In this figure, the increase of the perturbative uncertainty when going to higher slepton masses is clearly visible. For $\ptcut=25 \gev$ the relative NLL uncertainty increases from $24 \%$ at $m_{\slep}=300 \gev$ to $38\%$ at $m_{\slep}=1000 \gev$.
Going from NLL to NLL$'$+NLO, we observe a significant improvement. The NLL$'$+NLO uncertainty is roughly a factor of three to four smaller, and increases from $5.8 \%$ at $m_{\slep}=300 \gev$ to $11.2 \%$ at $m_{\slep}=1000 \gev$.
The corresponding results for $\slepR \slepR$ production are very similar. Finally, we note that it is certainly feasible if necessary to further reduce the perturbative uncertainties by going one order higher to NNLL$'$.

\section{Conclusions}
\label{sec:conclusions}

To maximize their sensitivity, several LHC searches for new physics require a specific number of signal jets and veto additional jets with transverse momentum above a certain value $\ptcut$, typically around 20-50 GeV. This jet veto introduces large logarithms of $\ptcut$ over the scale of new physics in the cross section, which requires resummation to obtain the best possible predictions.

We have presented the first predictions of a SUSY cross section including the higher-order resummation of jet-veto logarithms. Focusing on slepton (selectron and smuon) production, where a 0-jet sample is selected, we carry out resummation at NLL$'$ order and match our resummed results to the NLO cross section. Here we utilize the SCET framework for jet veto resummation developed in Higgs production.
Our analysis can also be extended to other new physics processes, including those with final-state jets (e.g. stop/sbottom production), which however also pose additional challenges due to the additional scales involved. 

A central aspect of our study is a systematic and thorough assessment of the theory uncertainty associated with the jet veto,
which we estimate using resummation profile scales. At the low resummation order provided by parton showers,
this uncertainty is substantial and not accounted for in current exclusion limits quoted by ATLAS and CMS. 
The higher-order resummed predictions provide much improved precision and will thus benefit the interpretation of the experimental observations. One possibility to easily utilize these (and future) theoretical improvements, is for the experimental analyses to also provide results for $\sigma^{95}_{0,\rm vis}$.

At the 13 TeV LHC run II the slepton mass reach is expected to increase up to $\sim 500\gev$ and beyond with $100 {\rm \;fb}^{-1}$ (see e.g.~refs.~\cite{Eckel:2014dza,Gershtein:2013iqa}). Our results show that the impact of the jet veto increases further at higher slepton masses, as expected. We provide precise resummed predictions for the 0-jet slepton cross sections at 13 TeV up to slepton masses of 1 TeV. Our predictions are available upon request. We hope that these results will allow the experimental analyses to continue relying on and benefiting from jet vetoes in optimizing the experimental sensitivity to new physics.
And once discovered, accurate theory predictions will be important to reveal the nature of any new particle.


\begin{acknowledgments}
We thank Kazuki Sakurai for helpful discussions and all the {\tt ATOM} authors for providing us with a version of their code. We also thank Stefan Liebler and Piotr Pietrulewicz for comments on the manuscript.
This work was supported by the German Science Foundation (DFG) through the Emmy-Noether Grant No. TA 867/1-1, by the Netherlands Organization for Scientific Research (NWO) through a VENI grant, and the D-ITP consortium, a program of the NWO that is funded by the Dutch Ministry of Education, Culture and Science (OCW).
\end{acknowledgments}

\appendix

\section{Fixed-order ingredients}
\label{app:FO}

\subsection{Hard function}
\label{app:hard}

The hard function consists of the Born cross section $\sigma_B$ and virtual corrections,
\begin{align}
H_{q\bar{q}} (Q^2, m_\text{SUSY}, \mu) 
= \si_B (1 + V)
\,.\end{align}
The Born cross section for slepton production is (see \fig{LOSlepton}) 
\begin{align} \label{eq:born}
 \sigma_B = \frac{\alpha_{\rm em}^2 \pi}{9 Q^2} \frac{1}{E_{\rm cm}^2}\, 
 \bigg(1- \frac{4m^2_{\tilde{\ell}_s}}{Q^2}\bigg)^{3/2} 
  h_{\tilde{\ell}_{s} \tilde{\ell}_{s}}
\end{align}
where the index $s=L,R$ labels the slepton state. The couplings enter in
\begin{align}
h_{\tilde{\ell}_{s} \tilde{\ell}_{s}}= 
Q_q^2 Q_\ell^2   
+ Q_q Q_\ell\, \frac{({g_q^-} + {g_q^+})({g_\ell^-}     \delta_{sL} + {g_\ell^+}    \delta_{sR} )}{1-m_Z^2/Q^2} 
+ \frac{({g_q^-}^2 + {g_q^+}^2)  ({g_\ell^-}^2 \delta_{sL} + {g_\ell^+}^2 \delta_{sR} )}{2 (1-m_Z^2/Q^2)^2} 
\,,\end{align}
where $Q_q$ and $Q_\ell$ are the electric charges, and $g_q^\pm, g_\ell^\pm$ are the couplings to the $Z$ boson
\begin{align}
g_f^- = \frac{I_f^3 - \sin^2 \theta_W\, Q_f}{\sin \theta_W\, \cos \theta_W} 
\,, \qquad 
g_f^+ = -\frac{\sin \theta_W\, Q_f}{\cos \theta_W}
\,,\end{align}

For the one-loop virtual corrections from QCD and SUSY-QCD, which are shown in \figs{NLOSleptonQCD}{NLOSleptonSUSY}, we get
\begin{align}
V &= \frac{\alpha_s(\mu) C_F}{4 \pi}\,  (V_{\rm QCD} +V_{\rm SUSY}) + h.c.
\nn \\ 
V_{\rm QCD} &= -\ln^2\Big(\frac{Q^2}{\mu^2}\Big) + 3 \ln\Big(\frac{Q^2}{\mu^2}\Big) - 8 + \frac{7 \pi^2}{6}
\nn \\
V_{\rm SUSY} &= 1 + \frac{2\,m_{\tilde g}^2 - 2 m_{\tilde q}^2}{Q^2} \big[ B_0 (Q^2, m_{\tilde q}^2, m_{\tilde q}^2) - B_0 (0, m_{\tilde g}^2,m_{\tilde q}^2)\big] + B_0 (Q^2, m_{\tilde q}^2, m_{\tilde q}^2) 
\nn \\ & \quad
+ 2 \frac{m_{\tilde g}^4+(Q^2-2 m_{\tilde q}^2)\,m_{\tilde g}^2 +m_{\tilde q}^4}{Q^2}\,
C_0 (0,0,Q^2,m_{\tilde q}^2,m_{\tilde g}^2,m_{\tilde q}^2)
\nn \\ & \quad
 - B_0 (0, m_{\tilde g}^2,m_{\tilde q}^2) + (m_{\tilde q}^2 - m_{\tilde g}^2) B'_0 (0,m_{\tilde g}^2,m_{\tilde q}^2) 
\,,\end{align}
where we have neglected squark mixing. 
This is in agreement with the expressions in refs.~\cite{Manohar:2003vb, Bauer:2003di, Djouadi:1999ht, Bozzi:2007qr}.
$B_0$ and $C_0$ are the scalar one-loop integrals, for which we use the {\tt LoopTools} conventions~\cite{Hahn:1998yk}.
Note that $V_{\rm SUSY}$ has no IR divergences in the full theory and hence does not have an explicit $\mu$ dependence and
therefore cannot change the anomalous dimensions of the SCET hard function for Drell-Yan.

Since we consider a simplified model with heavy squarks and gluinos, the SUSY-QCD corrections are much smaller than the QCD corrections. In our numerical results we choose $m_{\tilde g} = m_{\tilde q} = 4\TeV$, though the precise value in this region is irrelevant.

\subsection{Beam function}
\label{app:beam}

The (anti)quark beam function can be computed as a convolution of perturbative matching coefficients, $\cI_{qj}$, and the standard PDFs, $f_j$,
\begin{align}  \label{eq:beamFO}
B_{q} (p_T^{\rm cut}, x, \mu, \nu)
= \sum_j \int_x^1\! \frac{\df z}{z}\, \cI_{qj}(p_T^{\rm cut}, z, \mu, \nu)\, f_j\Bigl(\frac{x}{z}, \mu \Bigr)
\,.\end{align}
The matching coefficients expanded to NLO are
\begin{equation}
\cI_{qj} (p_T^{\rm cut}, z, \mu, \nu)
= \delta_{qj} \delta(1-z) + \frac{\als(\mu)}{4\pi}\, \cI_{qj}^{(1)} (p_T^{\rm cut}, z, \mu, \nu)+ \ord{\als^2}
\,.\end{equation}
The rapidity-renormalized $\ord{\alpha_s}$ matching coefficients were extracted from the calculations in ref.~\cite{Stewart:2013faa},
\begin{align}
\cI_{qq}^{(1)}(\ptcut, z, \mu, \nu)
&= 2C_F \biggl\{\ln \frac{\mu}{\ptcut}\,\Bigl[\Bigr(4\ln \frac{\nu}{Q} + 3\Bigr)\, \delta(1-z) - 2 P_{qq}(z)\Bigr] + I_{qq}(z) \biggr\}
\,, \nn \\
\cI_{qg}^{(1)}(\ptcut, z, \mu, \nu)
&= 2T_F \Bigl[-2 \ln \frac{\mu}{\ptcut}\, P_{qg}(z) + I_{qg}(z) \Bigr]
\,,\end{align}
with
\begin{align} 
P_{qq}(z) &= \biggl[\frac{\theta(1-z)}{1 - z}\biggr]_+(1 + z^2) + \frac{3}{2}\,\delta(1-z)
\nn \\
P_{qg}(z) &= \theta(1 - z)\bigl[(1 - z)^2 + z^2\bigr]
\nn \\
I_{qq}(z) &= 1 - z
\nn \\
I_{qg}(z) &= 2z(1 - z)
\,.\end{align}
These agree with the results in refs.~\cite{Liu:2012sz, Ritzmann:2014mka, Luebbert:2016itl}.


\subsection{Soft function}
\label{app:soft}

The NLO soft function is obtained from ref.~\cite{Stewart:2013faa} using Casimir scaling
\begin{align} 
\label{eq:softFO}
&S_{q\bar q}(p_T^{\rm cut}, \mu, \nu)
=
1 + \frac{\als(\mu)}{4\pi}\,
C_F\biggl[ 8\ln \frac{\mu}{\ptcut} \Bigl( \ln \frac{\mu}{\ptcut} - 2  \ln \frac{\nu}{\ptcut}\Bigr) - \frac{\pi^2}{3} \biggr]\,.
\end{align}

\subsection{Nonsingular contributions}
\label{app:nons}

\begin{figure}
\includegraphics[width=0.5\textwidth]{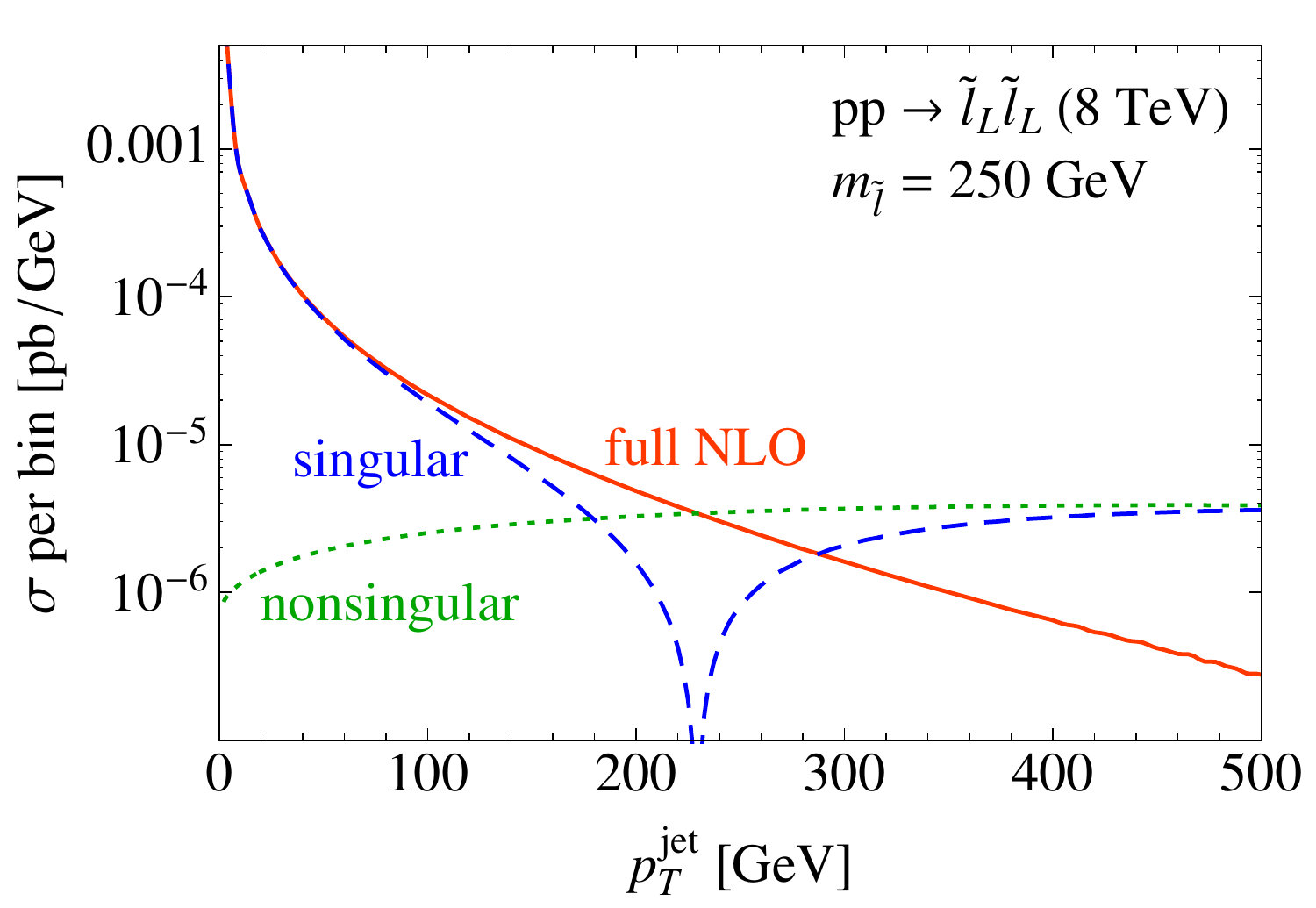}%
\hfill
\includegraphics[width=0.5\textwidth]{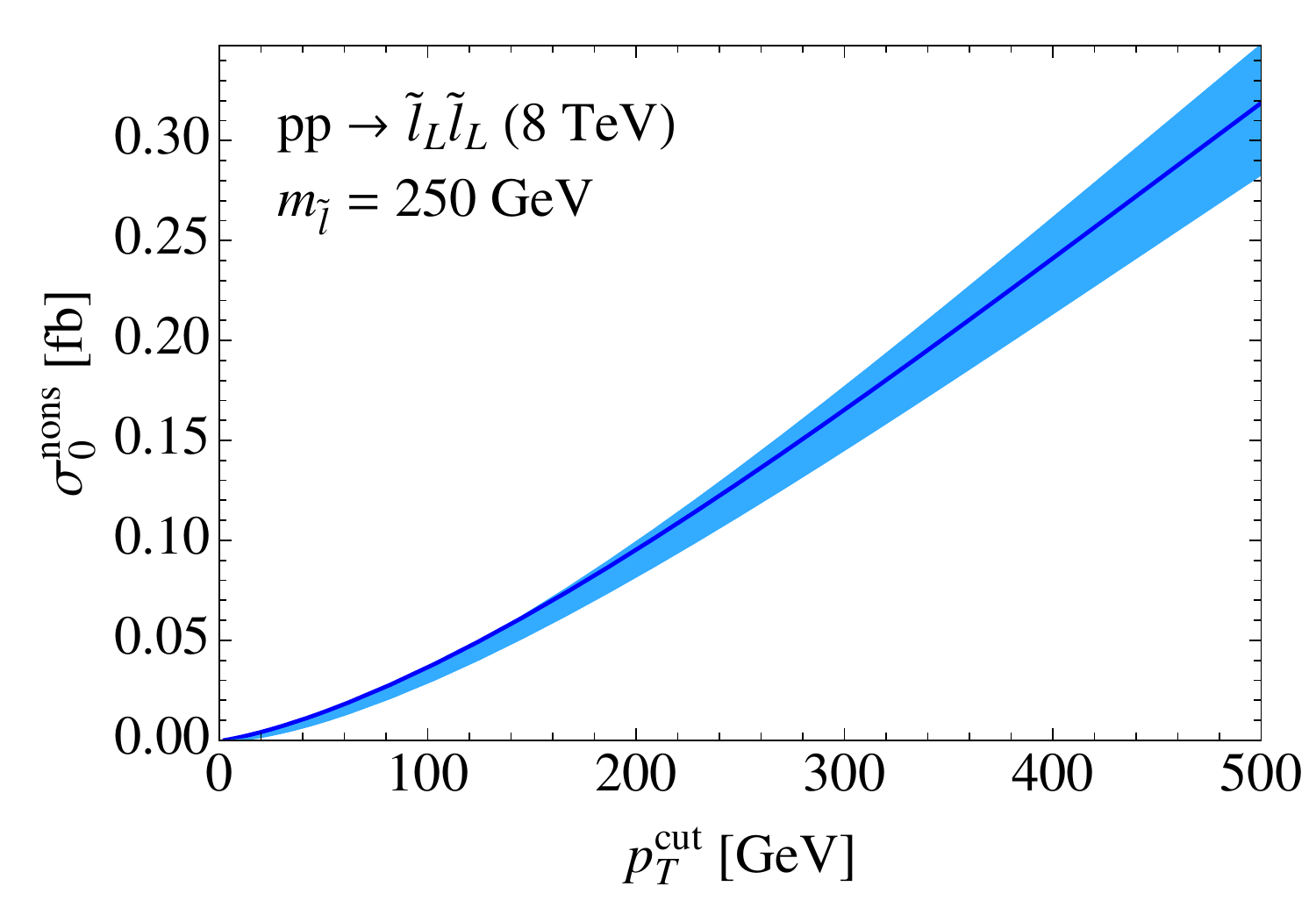}
\caption{Left: Singular (blue dashed) and nonsingular (green dotted) contributions to the full NLO (red solid) differential cross section for $\slepL \slepL$ production. Right: The (integrated) nonsingular cross section for $\slepL \slepL$ at NLO.}
\label{fig:singvsnonsing8}
\end{figure}

The fixed-order cross section can be split into a singular part and a nonsingular part,
\begin{align}
\sigma_0^{\rm FO}(\ptcut) &= \sigma_0^{\rm sing} (\ptcut) + \sigma_0^{\rm nons} (\ptcut)
\label{eq:nonsingular1}
\,,\end{align}
where we suppress the dependence on the SUSY masses for simplicity.
The logarithmically enhanced terms in the singular cross section, $\sigma_0^{\rm sing}(\ptcut)$, are contained in the resummed part in \eq{fac}. The nonsingular cross section, $\sigma_0^{\rm nons}(\ptcut)$, contains terms which scale as $\ord{p_T^\text{cut}/Q}$ and vanishes for $\ptcut \to 0$. In this section we discuss how to extract  $\sigma_0^{\rm nons}(\ptcut)$, which is essential to reproduce the correct fixed-order cross section for large $p_T^\cut$.

As suggested by \eq{nonsingular1}, the NLO nonsingular cross section can be extracted from the full NLO cross section and the NLO singular cross section. We achieve this using
\begin{align}
\sigma_0^{\rm nons}(\ptcut)
=  \int_{\eps \to 0}^{\ptcut}\!\df p_T^{\rm jet} \biggl( \frac{\df \sigma_0^{\rm FO} }{\df p_T^{\rm jet}} - \frac{\df \sigma_0^{\rm sing}}{\df p_T^{\rm jet}}\biggr)\,.
\end{align}
The left panel of \fig{singvsnonsing8} shows the NLO results for $\df \sigma_0^{\rm FO} / \df p_T^{\rm jet}$ (red solid), $\df \sigma_0^{\rm s} / \df p_T^{\rm jet}$ (blue dashed) and their difference $\df \sigma_0^{\rm nons} / \df p_T^{\rm jet}$ (green dotted).
We determine the NLO singular cross section, by setting all scales in the NLL$'$ result equal to $\mu_{\rm FO}$, thus switching off the resummation. The full NLO cross section, differential in $p_T^{\rm jet}$, is obtained by generating about 3 million events for $pp \to \slep \slep + j$ using {\tt Madgraph} 2.3.2~\cite{Alwall:2014hca} with a lower cutoff on $p_T^{\rm jet}$ of $0.2 \gev$.
For small $p_T^{\rm jet}$ a precise cancellation between large values of $\df \sigma_0^{\rm FO} / \df p_T^{\rm jet}$ and $\df \sigma_0^{\rm s} / \df p_T^{\rm jet}$ is needed to obtain a reliable result for the nonsingular cross section, see \fig{singvsnonsing8}.
This is achieved using a large number of Monte Carlo events and fitting the nonsingular to the functional form
\begin{align}
\frac{\df \sigma_0^{\rm nons}}{\df p_T^{\rm jet}} = a \ln \frac{p_T^{\rm jet}}{2 m_{\tilde \ell}}+b+c \frac{p_T^{\rm jet}}{2 m_{\tilde \ell}} \ln \frac{p_T^{\rm jet}}{2 m_{\tilde \ell}} + d \frac{p_T^{\rm jet}}{2 m_{\tilde \ell}}
\,,\end{align}
which has the correct leading behavior for the differential spectrum for $p_T^\jet\to 0$.
In this fit all points with $p_T^\text{\rm jet} < x$ are included, where the default is $x=2 m_{\tilde \ell}$. As an important cross check, we ensure that the fitted result is stable under varying $x$.
The left panel of \fig{singvsnonsing8} shows that for $p_T^{\rm jet} \gtrsim m_{\slep}$, the nonsingular contributions are of the same size as the singular contributions, requiring their inclusion to correctly reproduce the full fixed-order cross section.
Our final results for the NLO $\sigma_0^{\rm nons} (\ptcut)$ can be seen in the right panel of \fig{singvsnonsing8}. The band indicates the perturbative uncertainty, and is obtained by calculating the nonsingular terms three times, evaluating the ingredients at $\mu_{\rm FO} = m_{\slep}, 2 m_{\slep}$ and $4 m_{\slep}$. The nonsingular for right-handed slepton production is obtained in the same manner.

\section{RGE ingredients}
\label{app:RGE}

As explained in \sec{factorization}, the resummation of large logarithms is achieved in SCET by first evaluating the functions in the factorized cross section \eq{fac} at their natural virtuality ($\mu_H, \mu_B, \mu_S$) and rapidity ($\nu_B, \nu_S$) scales, and then RG evolving them to (arbitrary) common scales $\mu$ and $\nu$.
Writing this evolution out explicitly, \eq{fac} for inclusive slepton production becomes
\begin{align}
\sigma_0(p_T^{\rm cut}, m_\text{SUSY})
&= \!\int\! \df Q^2\, \df Y H_{q\bar{q}} (Q^2, m_\text{SUSY}, \mu_H)
\nn \\ & \quad \times
B_{q} (p_T^{\rm cut}, x_a, \mu_B, \nu_B) \, B_{\bar{q}} (p_T^{\rm cut}, x_b, \mu_B, \nu_B) \, S_{q\bar q}(p_T^{\rm cut}, \mu_S, \nu_S)
\nn \\ & \quad \times
U_0 (p_T^{\rm cut},  Q^2; \mu_H, \mu_B, \mu_S, \nu_B, \nu_S)
+ \sigma_0^\text{nons}(p_T^\cut, m_\text{SUSY})
\,.\end{align}
At NLL$'$ (NLL) order, we have to include the NLO (LO) results for the hard, beam and soft functions, given in \app{FO}. The evolution factor $U_0$ is given by the product of the individual evolution factors that evolve each of the functions from their natural scale to the common scales $\mu$ and $\nu$,
\begin{align} \label{eq:evo_kernel}
U_0 (p_T^{\rm cut},  Q^2; \mu_H, \mu_B, \mu_S, \nu_B, \nu_S)
&= \biggl\lvert
\exp\biggl[ \int_{\mu_H}^{\mu}\! \frac{\df\mu'}{\mu'}\, \gamma_H^q (Q^2, \mu') \biggr]
\biggr\rvert^2
\nn \\ & \quad \times
\exp \biggl[\int_{\mu_B}^\mu\! \frac{\df\mu'}{\mu'}\, 2\,\gamma_B^q(Q, \mu', \nu) \biggr]
\exp \biggl[ \int_{\mu_S}^\mu\! \frac{\df\mu'}{\mu'}\, \gamma_S^q(\mu', \nu) \biggr]
\nn \\ & \quad \times
\exp \biggl[ \ln\frac{\nu_B}{\nu} \gamma_\nu^q(\ptcut, \mu_B)  + \ln\frac{\nu}{\nu_S} \gamma_\nu^q(\ptcut, \mu_S) \biggr]
\,.\end{align}
The anomalous dimensions entering here are collected in the next subsection.
Note that due to RGE consistency the dependence on the arbitrary scales $\mu$ and $\nu$ exactly cancels between the different
factors in \eq{evo_kernel}.

\subsection{Anomalous dimensions}
\label{app:anom}

The anomalous dimension of the hard, beam, and soft functions that enter in the evolution kernel in \eq{evo_kernel} have the following general structure~\cite{Manohar:2003vb, Tackmann:2012bt}
\begin{align} \label{eq:anom_dim}
\gamma_H^q (Q^2, \mu) 
&= 
\Gcusp [\als(\mu)] \ln \frac{Q^2}{\mu^2} + \gamma_H^q [\als(\mu)]
\,, \nn \\
\gamma_B^q(Q,\mu, \nu)
&= 2 \Gcusp [\als(\mu)] \ln\frac{\nu}{Q} + \gamma_B^q[\als(\mu)]
\,, \nn \\
\gamma_S^q(\mu, \nu)
&= 4\Gcusp [\als(\mu)] \ln \frac{\mu}{\nu} + \gamma_S^q[\als(\mu)]
\,, \nn \\
\gamma_\nu^q(\ptcut, \mu) &= -4\eta_\Gamma^q(\ptcut,\mu) + \gamma_\nu^q[\als(\ptcut)]
\,,\end{align}
where the exact path-independence of the evolution in $(\mu,\nu)$ space~\cite{Chiu:2012ir} is ensured by
\begin{equation}
\eta^q_{\Gamma}(\mu_0, \mu)
= \int_{\mu_0}^\mu\! \frac{\df\mu'}{\mu'}\, \Gcusp[\als(\mu')]
\,.\end{equation}
The exact $\mu$ independence of the cross section is equivalent to the RG consistency relation
\begin{align} \label{eq:anom_consistency}
2 \gamma_H^q (Q^2, \mu) + 2 \gamma_B^q(Q,\mu, \nu) + \gamma_S^q(\mu, \nu) = 0\,.
\end{align}

We give the cusp and noncusp anomalous dimensions in terms of an expansion in $\alpha_s$,
\begin{align}
\label{eq:anodevo}
\Gamma_\cusp^q(\alpha_s) = \sum_{n=0}^\infty \Gamma_n^q \Bigl(\frac{\alpha_s}{4\pi}\Bigr)^{n+1}
\,, \qquad
\gamma_X^q(\alpha_s) = \sum_{n=0}^\infty \gamma_{X\,n}^q \Bigl(\frac{\alpha_s}{4\pi}\Bigr)^{n+1}
\,.\end{align}
At NLL (and NLL$'$) we require the one-loop noncusp anomalous dimensions $\ga_{X\,0}^q$ and the
two-loop cusp anomalous dimension, $\Ga_0^q$, $\Ga_1^q$, as well as the two-loop running for $\alpha_s$.
At NNLL we would need each at one order higher, which we also give below.

The coefficients for the cusp anomalous dimension are~\cite{Korchemsky:1987wg, Moch:2004pa}
\begin{align} 
\Gamma_0^q &= 4 C_F
\,,\nn\\
\Gamma_1^q &= 4 C_F \Bigl[\Bigl( \frac{67}{9} -\frac{\pi^2}{3} \Bigr)\,C_A  -
   \frac{20}{9}\,T_F\, n_f \Bigr]
\,,\nn\\
\Gamma_2^q &= 4 C_F \Bigl[
\Bigl(\frac{245}{6} -\frac{134 \pi^2}{27} + \frac{11 \pi ^4}{45}
  + \frac{22 \zeta_3}{3}\Bigr)C_A^2
  + \Bigl(- \frac{418}{27} + \frac{40 \pi^2}{27}  - \frac{56 \zeta_3}{3} \Bigr)C_A\, T_F\,n_f
\nn\\* & \hspace{8ex}
  + \Bigl(- \frac{55}{3} + 16 \zeta_3 \Bigr) C_F\, T_F\,n_f
  - \frac{16}{27}\,T_F^2\, n_f^2 \Bigr]
\,.\end{align}
The hard noncusp anomalous dimension is those of the quark form factor~\cite{Kramer:1986sg, Matsuura:1988sm}. The noncusp anomalous dimension coefficients for the soft function and rapidity evolution follow from ref.~\cite{Stewart:2013faa} using Casimir scaling, and those for the beam function then follow from the consistency relation in \eq{anom_consistency}. This leads to
\begin{align} 
\gamma^q_{H\,0} &= -6 C_F
\,,\nn\\
\gamma^q_{H\,1}
&= - C_F \Bigl[
  \Bigl(\frac{82}{9} - 52 \zeta_3\Bigr) C_A
+ (3 - 4 \pi^2 + 48 \zeta_3) C_F
+ \Bigl(\frac{65}{9} + \pi^2 \Bigr) \beta_0 \Bigr]
\,,\nn \\[1ex]
\gamma^q_{B\,0} &= 6 C_F
\,,\nn\\
\gamma^q_{B\,1} &= C_F  \biggl[  ( 3 - 4 \pi^2 + 48 \zeta_3 ) C_F + (-14 + 16 (1+\pi^2) \ln 2 - 96 \zeta_3) C_A
\nn\\
&\phantom{= C_F  a}
+ \Bigl( \frac{19}{3} - \frac{4}{3} \pi^2 + \frac{80}{3} \ln 2 \Bigr) \beta_0  \biggr]
\,, \nn \\[1ex]
\gamma^q_{S\,0} &= 0
\,,\nn\\
\gamma^q_{S\,1} &= 8 C_F \biggl[ \Bigl(\frac{52}{9} - 4 (1+\pi^2) \ln 2 + 11 \zeta_3\Bigr) C_A
+ \Bigl(\frac{2}{9} + \frac{7 \pi^2}{12} - \frac{20}{3} \ln 2 \Bigr) \beta_0 \biggr]
\,, \nn \\[1ex]
\gamma_{\nu\,0}^q &= 0
\,,\nn \\
\gamma_{\nu\,1}^q &= -16 C_F \biggl[ \Bigl( \frac{17}{9} - (1+\pi^2) \ln 2 + \zeta_3\Bigr) C_A + \Bigl( \frac{4}{9} + \frac{\pi^2}{12} -\frac{5}{3} \ln 2\Bigr) \beta_0 \biggr] + C_2 (R)
\,,\end{align}
where $C_2 (R) = 16 C_F C_A (-2.49 \ln R^2 - 0.49 ) + {\cal O}(R^2)$ denotes the clustering correction from the jet algorithm~\cite{Stewart:2013faa}. For completeness, in our convention we have
\begin{align}
\beta_0 &= \frac{11}{3}\,C_A -\frac{4}{3}\,T_F\,n_f
\,,\qquad
C_A = N_c
\,, \qquad
 C_F = \frac{N_c^2 - 1}{2N_c} = \frac43
 \,, \qquad
 T_F = \frac12
\,,\end{align}
where $N_c = 3$ is the number of colors and $n_f = 5$ is the number of active quark flavors.

\subsection{Profiles scales}
\label{app:profiles}

In this appendix we give the expressions for the scales $\mu_H, \mu_B, \mu_S$ and $\nu_B, \nu_S$ employed for our central value and uncertainty estimate. A discussion of our $p_T$-dependent profile scales is given in \sec{unc}, and includes plots and our procedure for estimating the perturbative uncertainty.

At small values of $\ptcut$ the full NLO cross section is governed by the singular cross section containing the logarithmic terms which need to be resummed; see the left panel of \fig{singvsnonsing8} and its discussion.
From the anomalous dimensions in \eq{anom_dim} we can read off the canonical scales already given in \eq{natural} for which the logarithms in the functions are minimized,
\begin{align} \label{eq:canonical}
\mu_H &= 2 m_{\slep} \sim Q
\,, \nn \\
\mu_B &= \ptcut \,, \qquad \nu_B= 2 m_{\slep} \sim Q
\,, \nn \\
\mu_S &= \ptcut \,, \qquad \nu_S= \ptcut
\,.\end{align}  
These are the appropriate scale choices in the resummation region.

At large values $\ptcut \sim Q$, singular and nonsingular contributions are of similar size and there are large cancellations between them. This can be observed in the left panel of \fig{singvsnonsing8}, where for $\ptcut \gtrsim 300 \gev$ the singular and nonsingular contributions have larger magnitudes (and opposite signs) than the full result.
To reproduce this cancellation and thus the fixed-order result, resummation must be turned off at this point. This is achieved by evaluating all functions in the factorized cross section at a common fixed-order scale
\begin{align}
\mu_H = \mu_B = \mu_S  = \nu_B = \nu_S = \mu_{\rm FO} = 2 m_{\slep}
\,,\end{align}
which is also the scale used for the nonsingular corrections. The value $\mu_{\rm FO} = 2 m_{\slep} \sim Q$ is chosen to agree with the value of $\mu_H$ used at small $\ptcut$. In the intermediate region, both resummation and fixed order terms are relevant. In this region, the scales are chosen to smoothly interpolate between the resummation region at small $\ptcut$ values and the fixed-order region at large $\ptcut$ values.

We follow ref.~\cite{Stewart:2013faa} and choose our (central) profile scales according to
\begin{align} 
\mu_H &= \nu_B = \mu_{\rm FO}
\,,\nn\\
\mu_B &= \mu_S = \nu_S = \mu_{\rm FO} \times f_{\rm run} \bigl(\ptcut/(2 m_{\slep})\bigr)
\,,
\end{align}
with 
\begin{align} \label{eq:profile}
f_{\rm run} (x) =
\begin{mcases}[ll@{\ }l]
x_0 \bigl[ 1 + (x / x_0)^2/4 \bigr] & x \le 2x_0 & \text{ nonperturbative region}
\\
x & 2x_0 \le x \le x_1 & \text{ resummation region}
\\
x + \dfrac{(2 - x_2 - x_3) (x - x_1)^2}{2(x_2 - x_1) (x_3 - x_1)} & x_1 \le x \le x_2  & \text{ transition from resummation}
\vspace{1ex} \\
1 - \dfrac{(2 - x_1 - x_2) (x - x_3)^2}{2(x_3 - x_1) (x_3 - x_2)} & x_2 \le x \le x_3 & \text{ transition to fixed order}
\\
1 & x_3 \le x & \text{ fixed-order region}
\end{mcases}
\end{align}
The values for $x_1$, $x_2$, $x_3$ determine where the transition from resummation to fixed-order region happens. They are chosen as
\begin{equation}
\{x_1, x_2, x_3\} = \{0.15 ,\, 0.4 ,\, 0.65 \}
\end{equation}
by considering the relative size of the singular and nonsingular terms in \fig{singvsnonsing8}.
Below $x_1$ we have exact canonical running, \eq{canonical}, while above $x_3$ the resummation is fully turned off.
For $2m_{\slep} = 500 \gev$ this corresponds to \{75,\,200,\,325\} GeV.
In addition we choose $x_0 = 2.5 \GeV / \mu_{\rm FO}$.
The resulting central scales are shown as solid blue ($\mu_H, \nu_B$) and red ($\mu_B, \mu_S, \nu_S$) lines in \fig{profiles}.

To estimate the perturbative uncertainties in the resummed prediction, variations of the profile scales are considered, as discussed in \sec{unc}. Here we very briefly summarize the variations; more details on their derivation can be found in ref.~\cite{Stewart:2013faa}.
The set of variations $V_{\mu}$ determining $\Delta_{\mu 0}$ has 14 profile scale variations, which are all possible combinations of
\begin{enumerate}
\item an overall up and down variation of the fixed-order scale $\mu_\text{FO}$ by factors of 2 and 1/2,
\item four variations for the transition points $x_1,\,x_2,\,x_3$
\begin{align}
\{x_1, x_2, x_3\} \; : \; &\{0.1, 0.3, 0.5\} \,, \{0.2, 0.5, 0.8\} \,,
\{0.04, 0.4, 0.8\}\,, \{0.2, 0.35, 0.5\} \,.
\end{align}
\end{enumerate}

The set of variations $V_{\rm resum}$ of $\mu_B, \mu_S, \nu_B$, $\nu_S$ determining $\Delta_{\rm resum}$ are combinations of
\begin{align}
\mu_i^{\rm up} (\ptcut) &= \mu_i^{\rm central} (\ptcut) \times f_{\rm vary}\bigl(\ptcut/(2 m_{\slep})\bigr)\,,
\nn\\
\mu_i^{\rm down} (\ptcut) &= \mu_i^{\rm central} (\ptcut) \,/ \, f_{\rm vary}\bigl(\ptcut/(2 m_{\slep})\bigr)\,,
\nn\\
 \nu_i^{\rm up} (\ptcut)  &= \nu_i^{\rm central} (\ptcut) \times f_{\rm vary}\bigl(\ptcut/(2 m_{\slep})\bigr)\,,
 \nn \\
 \nu_i^{\rm down} (\ptcut) &= \nu_i^{\rm central} (\ptcut) \,/ \, f_{\rm vary}\bigl(\ptcut/(2 m_{\slep})\bigr)\,.
\end{align}
The multiplicative variation factor is defined as
\begin{align}
f_{\rm vary} (x) =
\begin{cases}
2(1 - x^2 / x_3^2) & 0 \le x \le x_3 / 2 \,, \\
1 + 2(1 - x/x_3)^2 & x_3 / 2 \le x \le x_3 \,, \\
1 & x_3 \le x \,,
\end{cases}
\end{align}
which approaches a factor of 2 for $\ptcut\to 0$ and turns off for $x\to x_3$.
Out of the $80$ possible combinations of variations, all combinations leading to arguments of logarithms which are more then a factor of 2 different from their central values are not considered. This leaves a total of 35 profile scale variations in $V_{\rm resum}$.

\bibliographystyle{jhep}
\bibliography{refs}


\end{document}